\documentclass[aps, prd, showkeys, showpacs]{revtex4}

\usepackage[small]{caption}
\usepackage{amsmath,amsfonts,amscd,amssymb,epsf, epsfig}\usepackage{graphicx}

\newcommand{\Z}{{\mathbb{Z}}}

\newcommand{\R}{\mathbb{R}}

\newcommand{\pa}{\partial}

\newcommand{\vep}{\varepsilon}

\begin{document}

 \title{ Casimir Piston of Real Materials and its Application to Multi-Layer Models}
 \author{L.P. Teo}
 \address{Department of Applied Mathematics, Faculty of Engineering, University of Nottingham Malaysia Campus, Jalan Broga, 43500, Semenyih, Selangor Darul Ehsan, Malysia. }\email{LeePeng.Teo@nottingham.edu.my}

\begin{abstract}
In this article, we derive the formula for the Casimir force acting on a piston made of real material moving inside a perfectly conducting rectangular box. It is shown that by taking suitable limits, one recovers the formula for the Casimir force acting on a perfectly conducting piston or an infinitely permeable piston.     Lifshitz formula for finite temperature Casimir force acting on parallel plates made of real materials is re-derived by considering the five-layer model in the context of piston approach. It is observed that the divergences of the Casimir force will only cancel under certain conditions, for example, when the regions separated by the plates are filled with   media of the same refractive index.
\end{abstract}

 \maketitle
 \section{Introduction}Casimir effect \cite{16} has wide applications in different areas of physics such as quantum field theory, gravitation and cosmology, atomic physics, condensed matter, nanotechnology and mathematical physics (see e.g. \cite{17, 18, 19, 20, 21, 22, 23, 24, 25, 26, 27}). According to Casimir, there is an attractive force
 \begin{equation}\label{eq8_26_1}F_{\text{Cas}}= -\frac{\pi^2 \hbar c A}{240 a^4}\end{equation} acting between two perfectly conducting plates with area $A$ and separation distance $a$. Naively the Casimir energy due to a quantum field is the sum of the zero point energies of the quantum field:
\begin{equation}\label{eq8_26_2}
E_{\text{Cas}} =\frac{\hbar}{2}\sum \omega,
\end{equation}where  the sum runs through all eigenfrequencies $\omega$ of the field. Generically, this sum is an infinite sum, and Casimir is the first one to extract a finite force \eqref{eq8_26_1} acting between two parallel perfectly conducting plates. Since then, various methods have been proposed for the regularization of the Casimir energy \eqref{eq8_26_2}.

 Casimir force is a pure quantum effect which can be considered as the macroscopic manifestation of the London-van der Waals force between atoms and molecules. Lifshitz \cite{3,28, 29} has laid a theoretical foundation for the description of both the London-van der Waals force and the Casimir force. The celebrated Lifshitz formula expresses the finite temperature Casimir force between two plane parallel plates as some functionals of the electric permittivities of the plate materials. This formula has been re-derived using different approaches \cite{17, 18, 19, 20, 21, 22, 23, 24, 25, 26, 27}, and is the main tool for the theoretical investigation of the Casimir effect on real materials at nonzero temperature. The original Lifshitz formula models the two parallel plates as two semi-infinite slabs which have infinite thickness. This approach is justified if the skin-depth of the material of the two plates is much smaller than the thickness of the plates. A refinement of the Lifshitz formula where the finite thickness of the plates is taken into consideration has been derived and usually related to the consideration of the multi-layer models \cite{new1, 5,9,10,11,12,13,54}.

Recently the issue of   divergences in Casimir effect was discussed extensively in \cite{new2,new3, new4, new5}. In connection to the application of Casimir effect in quantum field theory, this has been pointed out in \cite{30,31}. Recently, this issue was reconsidered in \cite{32,33,34,35} where it was argued that surface divergence terms in the Casimir energy cannot be removed by any renormalization of the physical parameters. Although the  arguments of \cite{32, 33, 34, 35}  have been refuted in the works \cite{36, 37}, one still need to be cautious in regularizing the Casimir energy. In \cite{38}, Cavalcanti introduced a geometric setup called piston and showed that the Casimir force acting on a piston is free of divergences, due to the cancelations of the divergences from the two regions separated by the piston. By taking certain limits, one can recover the Casimir force acting on a pair of parallel plates. Therefore, the piston approach provides an interpretation for the formula of the finite Casimir force acting between parallel plates. Since then, Casimir effect in the piston setup has attracted considerable interest \cite{39,40,7,41,42,43,6,8,44}. It has also been used to explore the Casimir effect in the presence of extra dimensions \cite{45,46,47,48,49,50,51,52}. Nevertheless, these works either considered scalar fields with Dirichlet or Neumann boundary conditions or more general Robin boundary conditions, or electromagnetic fields with perfect electric conductor conditions or perfect magnetic conductor conditions. In these cases, the piston becomes a barrier for the quantum fields to penetrate from one region to the other, and therefore one can compute the Casimir energies in the two regions separated by the piston independently. The divergent part of the Casimir energies in the two regions would give rise to Casimir force of opposite direction on the piston and therefore cancel each other.

Due to the success of the piston approach in the parallel plate scenario, one would expect a similar approach can be used to derive the Lifshitz formula for the Casimir effect on real materials. As a matter of fact, piston type approach has been employed by Schram \cite{53} in 1973 as a regularization scheme to derive the Lifshitz formula at zero temperature. The generalization to include the thermal effect is still lacking. Moreover, in the work of Schram, some subtraction scheme was employed to get rid of the divergences, instead of observing an automatic cancelation of divergence as in the recent works on pistons. In this work, we re-derive the Lifshitz formula using the piston approach. In fact, we consider a more general setup. First we derive the Casimir energy for a three-layer model, i.e., three parallel layers of different materials    with electric permittivities $\vep_1,\vep_2,\vep_3$ and magnetic permeabilities $\mu_1,\mu_2,\mu_3$. Using the piston approach, we first embed the layers   inside a large rectangular box which has perfectly conducting walls  serving as impenetrable barriers to  the electromagnetic field. The finite temperature Casimir energy for the electromagnetic field inside the box is computed using mode sum approach. An exponential cut-off is used to regularize the Casimir energy. The result is written as a divergent part (which goes to infinity if the cut-off parameter approaches zero) plus a finite interaction term (which is independent of the cut-off parameter). Treating the second layer as a non-deformable piston, we compute the Casimir force acting on a piston made of real material by differentiating with respect to the distance between the piston and one end of the rectangular box. The limit where the other end of the rectangular box is brought to infinity gives the Casimir force acting between a perfectly conducting plate and a plate made of real material. It is shown that the divergent part of the Casimir force will only be canceled under certain conditions, for example, when the regions separated by the piston or the plates are filled with   media  of the same refractive index $n=c\sqrt{\vep\mu}$. In general, one cannot ignore the cut-off parameter.

To derive  the Casimir force acting on two parallel plates both made with real materials, it is necessary to consider the five-layer model. Using the same approach as the three-layer model, we derive the Casimir energy and the Casimir force for the five-layer model. In the limit where both ends of the confining rectangular box go to infinity, we obtain  the Casimir force acting on two parallel plates, both made with real materials. The limit where the thickness of the two plates are infinite recovers the celebrated Lifshitz formula. As in the case of the three-layer model, we observe that the divergences of the Casimir force would cancel each other if the three regions separated by the two plates are filled with   media of the same refractive index.

Finally, we would like to remark  that in this work, the piston approach does not really refer to the piston, but rather the finite rectangular box with perfectly conducting walls that serve to confine the field into a finite region. The advantage of this approach is that in solving for the field modes, one do not have to use imaginary energy modes as in some of the previous works. This bridge the gap between the derivation of the Casimir force for real materials and the derivation of the Casimir force for ideal metals.

\section{Casimir energy inside a piston system}\label{s2}

 \begin{figure}
\epsfxsize=0.4\linewidth \epsffile{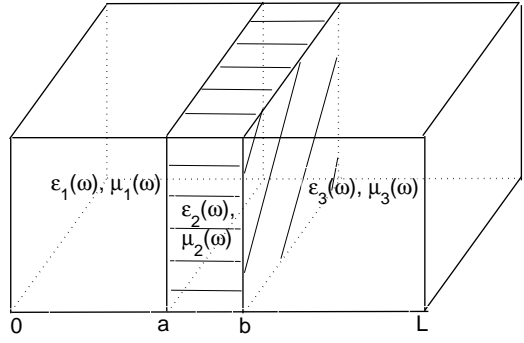} \caption{\label{f1}A   piston with electric permittivity $\vep_2$ and magnetic permeability $\mu_2$ diving the rectangular box into two regions, which are filled with materials having electric permittivities $\vep_1,\vep_3$ and magnetic permeabilities $\mu_1,\mu_3$. }\end{figure}
We consider a  piston with thickness $d_2=b-a$ dividing a rectangular box of dimension $L\times L_2\times L_3$ (see FIG. \ref{f1})   into two regions filled with materials with electric permittivities and magnetic permeabilities $\vep_1,   \mu_1$ and $  \vep_3, \mu_3$ respectively. The  piston is assumed to have electric permittivity  $\vep_2$ and magnetic permeability   $\mu_2$   and the walls of the rectangular box is assumed to be perfectly conducting.   Maxwell's equations take the usual form
\begin{equation}\label{eq3_26_1}\begin{split}
\nabla . \textbf{D}  =&\rho_f,\hspace{1cm}\nabla\times \textbf{E} +\frac{\pa \textbf{B}}{\pa t} =0,\\
\nabla . \textbf{B} =& 0,\hspace{1.2cm}
\nabla\times \textbf{H}-\frac{\pa\textbf{D}}{\pa t}=\textbf{J}_f,
\end{split}
\end{equation}where $\textbf{D}=\vep\textbf{E}$ and $\textbf{B}=\mu\textbf{H}$. As usual, we introduce the potential $\Phi$ and $\textbf{A}$ such that
\begin{equation*}\begin{split}
\textbf{E}=& -\frac{\pa \textbf{A}}{\pa t}-\nabla \Phi,\hspace{1cm}
\textbf{B}=\nabla\times\textbf{A},
\end{split}
\end{equation*}and impose the radiation gauge
\begin{equation*}
 \nabla .\textbf{A}=0.
\end{equation*}In the absence of free charges and free current, i.e., $\rho_f=0$ and $\textbf{J}_f=\textbf{0}$, we can impose the additional condition $\Phi=0$. Then the first three equations in \eqref{eq3_26_1} are automatically satisfied and the fourth equation give
\begin{equation}\label{eq3_26_2}
\left(\vep\mu\frac{\pa^2}{\pa t^2}-\nabla^2 \right)\textbf{A}=0.
\end{equation}For the boundary conditions, the continuity of $\textbf{D}\,.\,\textbf{n}, \textbf{E}\times \mathbf{n}, \textbf{B}\,.\,\mathbf{n}$ and $\textbf{H}\times\mathbf{n}$   across the boundaries implies that    $\vep E_{1}$, $E_2$, $E_3$ and $B_1$, $\frac{1}{\mu}B_2$, $\frac{1}{\mu}B_3$ must be continuous in the $x_1$ direction. On the other hand, the perfectly conducting conditions on the boundary of the rectangular box imply that on the boundaries $x_1=0$ and $x_1=L$, $E_2=E_3=0$ and $B_1=0$. On the boundaries $x_2=0$ and $x_2=L_2$, $E_1=E_3=0$ and $B_2=0$. On the boundaries $x_3=0$ and $x_3=L_3$, $E_1=E_2=0$ and $B_3=0$.

A solution of equation \eqref{eq3_26_2}  satisfying the gauge and the boundary conditions can be written in the form
\begin{equation*}
\begin{split}
A_1=&\pi \left(\left[\frac{k_2}{L_2}\right]^2+\left[\frac{k_3}{L_3}\right]^2\right) \phi(x_1) \sin \frac{\pi k_2x_2}{L_2}\sin\frac{\pi k_3x_3}{L_3}e^{-i\omega t},\\
A_2=&\frac{k_2}{L_2}\frac{\pa\phi(x_1)}{\pa x_1} \cos\frac{\pi k_2x_2}{L_2}\sin\frac{\pi k_3x_3}{L_3}e^{-i\omega t},\hspace{2cm}(k_2, k_3)\in \mathcal{S}_{\text{TM}}:= \mathbb{N}^2,\\
A_3=&\frac{k_3}{L_3}\frac{\pa\phi(x_1)}{\pa x_1} \sin\frac{\pi k_2x_2}{L_2}\cos\frac{\pi k_3x_3}{L_3}e^{-i\omega t},
\end{split}
\end{equation*}for TM modes, and
\begin{equation*}
\begin{split}
A_1=&0,\\
A_2=& \frac{k_3}{L_3} \psi(x_1) \cos\frac{\pi k_2x_2}{L_2}\sin\frac{\pi k_3x_3}{L_3}e^{-i\omega t},\hspace{2cm}(k_2, k_3)\in \mathcal{S}_{\text{TE}}: \mathbb{N}_0^2\setminus\{(0,0)\},\\
A_3=&-\frac{k_2}{L_2}\psi(x_1)\sin\frac{\pi k_2x_2}{L_2}\cos\frac{\pi k_3x_3}{L_3}e^{-i\omega t},
\end{split}
\end{equation*}for TE modes. Here $ \mathbb{N}_0=\mathbb{N}\cup\{0\}$. The functions $\vep \phi (x_1)$ and $\frac{\pa \phi(x_1)}{\pa x_1}$ are continuous, and $\phi(x_1)$ satisfies
\begin{equation*}
\frac{\pa^2\phi(x_1)}{\pa x_1^2}=-p(x_1)^2\phi(x_1),\hspace{1cm}\left.\frac{\pa \phi(x_1)}{\pa x_1}\right|_{x_1=0,L}=0,
\end{equation*}where $p(x_1)=p_j$ for $x_1$ in region $j$ whose  electric permittivity and magnetic permeability are $\varepsilon_j(\omega)$ and $\mu_j(\omega)$ respectively. The functions $\psi(x_1)$, $\frac{1}{\mu}\frac{\pa\psi(x_1)}{\pa x_1}$ are continuous, and $\psi(x_1)$ satisfies
\begin{equation*}
\frac{\pa^2\psi(x_1)}{\pa x_1^2}=-p(x_1)^2\psi(x_1), \hspace{1cm} \psi(0)=\psi(  L)=0.
\end{equation*}Finally, for $j=1,2,3$,
\begin{equation*}
\vep_j(\omega)\mu_j(\omega)\omega^2=p_j^2+\left[\frac{\pi k_2}{L_2}\right]^2+\left[\frac{\pi k_3}{L_3}\right]^2=p_j^2+\lambda_{\boldsymbol{k}}^2.
\end{equation*}
For the function $\phi(x_1)$, assume that
\begin{equation*}
\begin{split}
\phi(x_1)=\begin{cases}
\Lambda_1 e^{ip_1x_1}+\Lambda_2 e^{-ip_1x_1}, \hspace{1cm}&0\leq x_1\leq a,\\
\Lambda_3 e^{ip_2x_1}+\Lambda_4e^{-ip_2x_1}, & a\leq x_1\leq b,\\
\Lambda_5e^{ip_3x_1}+\Lambda_6e^{-ip_3x_1}, & b\leq x_1\leq L.
\end{cases}
\end{split}
\end{equation*}Then the boundary conditions give six equations for the six variables $\Lambda_1,\ldots,\Lambda_{6}$:
\begin{equation}\label{eq8_4_1}
\begin{split}
&\Lambda_1  -\Lambda_2  =0,\\
&\vep_1\left(\Lambda_1 e^{ip_1a}+\Lambda_2e^{-ip_1a}\right)=\vep_2\left( \Lambda_3e^{ip_2a}+\Lambda_4e^{-ip_2a}\right),\\
&ip_1\left(\Lambda_1 e^{ip_1a}-\Lambda_2e^{-ip_1a}\right)=ip_2\left( \Lambda_3e^{ip_2a}-\Lambda_4e^{-ip_2a}\right),\\
&\vep_2\left(\Lambda_3 e^{ip_2b}+\Lambda_4e^{-ip_2b}\right)=\vep_3\left( \Lambda_5e^{ip_3b}+\Lambda_6e^{-ip_3b}\right),\\
&ip_2\left(\Lambda_3 e^{ip_2b}-\Lambda_4e^{-ip_2b}\right)=ip_3\left( \Lambda_5e^{ip_3b}-\Lambda_6e^{-ip_3b}\right)\\
&\Lambda_5e^{ip_3L}-\Lambda_6e^{-ip_3L}=0.
\end{split}
\end{equation}These homogeneous system in $(\Lambda_1,\ldots, \Lambda_{6})^T$ has a nontrivial solution if and only if the determinant of the corresponding matrix is zero. However, we can also solve the system in the following way. The first equation and the last equation give
\begin{equation}\label{eq3_27_1}
\Lambda_2=\Lambda_1, \hspace{1cm} \Lambda_5 = \Lambda_{6}e^{-2ip_3L}.
\end{equation}The second and third, fourth and fifth equations give respectively
\begin{equation}\label{eq8_3_1}
\begin{split}
\begin{pmatrix} \Lambda_3\\ \Lambda_{4}\end{pmatrix} =&\frac{1}{2}\begin{pmatrix} r_{12 }^+ e^{i(p_1-p_2)a} & -r_{12 }^- e^{-i(p_1+p_2)a}\\
-r_{12 }^-e^{i(p_1+p_2)a}& r_{12 }^+ e^{-i(p_1-p_2)a}\end{pmatrix}\begin{pmatrix} \Lambda_1\\ \Lambda_2\end{pmatrix}=\frac{1}{2}\begin{pmatrix} r_{12 }^+ e^{i(p_1-p_2)a} - r_{12 }^- e^{-i(p_1+p_2)a}\\
-r_{12 }^-e^{i(p_1+p_2)a}+ r_{12 }^+ e^{-i(p_1-p_2)a}\end{pmatrix}\Lambda_1,
\\\begin{pmatrix} \Lambda_3\\ \Lambda_{4}\end{pmatrix} =&\frac{1}{2}\begin{pmatrix} r_{32}^+ e^{i(p_3-p_2)b} & -r_{32}^- e^{-i(p_3+p_2)b}\\
-r_{32}^-e^{i(p_3+p_2)b}& r_{32}^+ e^{-i(p_3-p_2)b}\end{pmatrix}\begin{pmatrix} \Lambda_5\\ \Lambda_6\end{pmatrix}=\frac{1}{2}\begin{pmatrix} r_{32}^+ e^{i(p_3-p_2)b} e^{-2ip_3L}- r_{32}^- e^{-i(p_3+p_2)b}\\
-r_{32}^-e^{i(p_3+p_2)b}e^{-2ip_3L}+ r_{32}^+ e^{-i(p_3-p_2)b}\end{pmatrix}\Lambda_6,
\end{split}
\end{equation}where $$r_{jk}^{\pm}=\frac{p_j}{p_k}\pm\frac{\vep_j}{\vep_k}.$$
Equating the  two equations in \eqref{eq8_3_1} give an equation of the form
\begin{equation*}
\begin{pmatrix} \alpha\\\beta\end{pmatrix}\Lambda_1=\begin{pmatrix} \gamma\\\delta \end{pmatrix}\Lambda_{6}
\end{equation*}which has a nontrivial solution if and only if $\alpha\delta -\beta\gamma=0$. In other words, the system \eqref{eq8_4_1} has a nontrivial solution if and only if
\begin{equation*}
\begin{split}
e^{-ip_3L}\Biggl\{ &\left(r_{12}^+e^{ip_1d_1}-r_{12}^-e^{-ip_1d_1}\right)\left(r_{32}^+e^{ip_3d_3}-r_{32}^-e^{-ip_3d_3}\right)e^{ip_2d_2}\\&-\left( r_{12}^+e^{-ip_1d_1}-r_{12}^-e^{ip_1d_1}\right)\left(r_{32}^+e^{-ip_3d_3}-r_{32}^-e^{ip_3d_3}\right)e^{-ip_2d_2}\Biggr\}=0,
\end{split}
\end{equation*}where $d_1=a$, $d_2=b-a$ and $d_3=L-b$.
Discarding the nonzero factor $e^{-ip_3L}$ and define
\begin{equation}\label{eq8_3_9}\begin{split}
F_{0;\text{TM}}(\omega, \boldsymbol{k}) =&   \left(r_{12}^+e^{ip_1d_1}-r_{12}^-e^{-ip_1d_1}\right)\left(r_{32}^+e^{ip_3d_3}-r_{32}^-e^{-ip_3d_3}\right)e^{ip_2d_2}\\&-\left( r_{12}^+e^{-ip_1d_1}-r_{12}^-e^{ip_1d_1}\right)\left(r_{32}^+e^{-ip_3d_3}-r_{32}^-e^{ip_3d_3}\right)e^{-ip_2d_2}, \hspace{0.5cm}\boldsymbol{k}=(k_2,k_3)\in \mathcal{S}_{\text{TM}}\end{split}
\end{equation} where for $j=1,2,3$,
\begin{equation*}
p_j(\omega,\boldsymbol{k}) =\sqrt{ \varepsilon_j(\omega)\mu_j(\omega)\omega^2-\lambda_{\boldsymbol{k}}^2}.
\end{equation*}The set of TM energy eigenmodes   is the union of the sets of real nonnegative zeros of $F_{0;\text{TM}}(\omega, \boldsymbol{k}), \boldsymbol{k}=(k_2,k_3)\in \mathcal{S}_{\text{TM}}$. For the TE modes, one can show analogously that the set of TE energy eigenmodes is the union of the sets of real nonnegative zeros of
\begin{equation}\label{eq8_3_10}\begin{split}
F_{0;\text{TE}}(\omega, \boldsymbol{k}) =&\left(s_{12}^+e^{ip_1d_1}-s_{12}^-e^{-ip_1d_1}\right)\left(s_{32}^+e^{ip_3d_3}-s_{32}^-e^{-ip_3d_3}\right)e^{ip_2d_2}\\&-\left( s_{12}^+e^{-ip_1d_1}-s_{12}^-e^{ip_1d_1}\right)\left(s_{32}^+e^{-ip_3d_3}-s_{32}^-e^{ip_3d_3}\right)e^{-ip_2d_2}, \hspace{1cm}\boldsymbol{k}=(k_2,k_3)\in \mathcal{S}_{\text{TE}},
\end{split}\end{equation} where
$$s_{jk}^{\pm}= 1\pm\frac{p_j\mu_k}{p_k\mu_j}.$$Compare \eqref{eq8_3_10} to \eqref{eq8_3_9}, we find that $F_{0;\text{TE}}(\omega, \boldsymbol{k}) $ is obtained from $F_{0;\text{TM}}(\omega, \boldsymbol{k})$ by replacing $r_{jk}^{\pm}$ by $s_{jk}^{\pm}$. In the following, we assume that all the zeros of $F_{0;\text{TM}}(\omega,\boldsymbol{k}),\boldsymbol{k}=(k_2,k_3)\in \mathcal{S}_{\text{TM}},$ and $F_{0;\text{TE}}(\omega,\boldsymbol{k}), \boldsymbol{k}=(k_2,k_3)\in \mathcal{S}_{\text{TE}},$ are real.

The finite temperature Casimir energy   inside the piston  is given by
\begin{equation}\label{eq8_3_4}
E_{\text{Cas}}( t_c) =\frac{\hbar}{2}\sum_{\text{modes}}\omega e^{-t_c\omega}+k_BT\sum_{\text{modes}}\log\left(1-\exp\left(-\frac{\hbar \omega}{k_B T}\right)\right),
\end{equation}where we have introduced a cut-off parameter $t_c$ to render the first sum finite. Using the generalized Abel-Plana summation formula \cite{1,2} (see Appendix \ref{a1}), we compute this finite temperature Casimir energy in Appendix \ref{a2}. Up to the terms that give non-trivial limits when $t_c\rightarrow 0^+$, the result is
\begin{equation}\label{eq8_4_5}
\begin{split}
E_{\text{Cas}}(t_c) =& \sum_{j=1}^3d_j\Xi_{j}(t_c)+\Delta E_{\text{Cas}},\end{split}
\end{equation}where\begin{equation}\label{eq8_3_13}
\begin{split}
\Xi_j(t_c) =\frac{1}{\pi}\left(\sum_{\boldsymbol{k}\in\mathcal{S}_{\text{TM}}}+\sum_{\boldsymbol{k}\in\mathcal{S}_{\text{TE}}}\right)\int_{p_j(\omega,\boldsymbol{k})\geq 0}\left\{
\frac{\hbar\omega}{2} e^{-t_c\omega}+k_BT\log\left(1-\exp\left(-\frac{\hbar\omega}{k_BT}\right)\right)\right\}dp_j(\omega,\boldsymbol{k}),
\end{split}
\end{equation}and
\begin{equation}\label{eq8_3_12}
\begin{split}
\Delta E_{\text{Cas}}=& \frac{k_BT}{2}\sum_{\boldsymbol{k}\in\mathcal{S}_{\text{TM}}}\sum_{l=-\infty}^{\infty}\log\Biggl\{\left(1-\Delta_{12}^{\text{TM}}(i\xi_l,\boldsymbol{k}) e^{-2q_1( \xi_l,\boldsymbol{k})d_1}\right)
\left(1-\Delta_{32}^{\text{TM}}(i\xi_l,\boldsymbol{k})e^{-2q_3( \xi_l,\boldsymbol{k})d_3}\right) \\& -\left(  e^{-2q_1( \xi_l,\boldsymbol{k})d_1}-\Delta_{12}^{\text{TM}}(i\xi_l,\boldsymbol{k}) \right)\left( e^{-2q_3( \xi_l,\boldsymbol{k})d_3} -\Delta_{32}^{\text{TM}}(i\xi_l,\boldsymbol{k})\right)e^{-2q_2( \xi_l,\boldsymbol{k})d_2}\Biggr\} +\left(\text{TM}\longrightarrow \text{TE}\right),
\end{split}
\end{equation}with
\begin{equation}\label{eq8_5_2}\begin{split}q_j(\xi,\boldsymbol{k})=&\sqrt{\varepsilon_j(i\xi)\mu_j(i\xi)\xi^2+\lambda_{\boldsymbol{k}}^2},\hspace{1cm}\xi_l=\frac{2\pi |l| k_BT}{\hbar},\\\Delta_{jk}^{\text{TM}}(i\xi,\boldsymbol{k})=&\frac{\vep_k(i\xi)q_j(\xi,\boldsymbol{k})-\vep_j(i\xi)q_k(\xi,\boldsymbol{k})}
{\vep_k(i\xi)q_j(\xi,\boldsymbol{k})+\vep_j(i\xi)q_k(\xi,\boldsymbol{k})},\hspace{0.5cm}\Delta_{jk}^{\text{TE}}(i\xi,\boldsymbol{k})=
\frac{\mu_j(i\xi)q_k(\xi,\boldsymbol{k})-\mu_k(i\xi)q_j(\xi,\boldsymbol{k})}{
\mu_j(i\xi)q_k(\xi,\boldsymbol{k})+\mu_k(i\xi)q_j(\xi,\boldsymbol{k})}.\end{split}\end{equation}
Notice that $\sum_{j=1}^3d_j\Xi_{j}(t_c)$ contains all the $t_c\rightarrow 0^+$ divergences. The term  $d_1\Xi_1(t_c)$ is independent of $  d_2, d_3$ and the parameters of media 2 and 3. It represents the self energy of media 1.  The term $\Xi_1(t_c)$ can be interpreted as the Casimir energy per unit length that would exists in the region between $x_1=0$ and $x_1=a$ if the boundaries at $x_1=0$ and $x_1=a$ were absent. Similar interpretation can be given to the terms $d_2\Xi_2(t_c)$ and $d_3\Xi_3(t_c)$. The term $\Delta E_{\text{Cas}}$ is called the interaction term. In the literatures on Casimir effect,   $\Delta E_{\text{Cas}}$ is usually regarded as the regularized Casimir energy by claiming that the term  $\sum_{j=1}^3d_j\Xi_{j}(t_c)$ should be subtracted away as the energy in the absence of boundary. This regularization scheme is a little superficial, since we have to consider the three regions that have different   properties separately. As explained in \cite{new2, new3, new4, new5}, one should not ignore the divergent terms, but should instead give a physical interpretation to the cut-off parameter $t_c$.  Nevertheless, in the special case $\Xi_1(t_c)=\Xi_2(t_c)=\Xi_3(t_c)$, then the term $d_1\Xi_1(t_c)+d_2\Xi_2(t_c)+d_3\Xi_3(t_c)=L\Xi_1(t_c)$ is independent of $d_1, d_2, d_3$, and the subtraction of this term  from the Casimir energy becomes  natural since it would not contribute to the Casimir force (see next section). $\Xi_1(t_c)=\Xi_2(t_c)=\Xi_3(t_c)$ holds when for example, the system is isorefractive, i.e., the refractive index $n(\omega)=c\sqrt{\vep(\omega)\mu(\omega)}$ is the same for the three materials.

The interaction term of the energy $\Delta E_{\text{Cas}}$ shows that in the high temperature limit, the leading term of the energy is linear in $T$, given by the sum of the $l=0$ terms. The zero temperature Casimir energy  is obtained by taking the limit $T\rightarrow 0$ in \eqref{eq8_3_13} and \eqref{eq8_3_12}, which gives
\begin{equation*}
E_{\text{Cas}}^{T=0}( t_c) =\frac{\hbar}{2\pi}\left(\sum_{\boldsymbol{k}\in\mathcal{S}_{\text{TM}}}+\sum_{\boldsymbol{k}\in\mathcal{S}_{\text{TE}}}\right)\int_{p_j(\omega,\boldsymbol{k})\geq 0}
 \omega  e^{-t_c\omega}dp_j(\omega,\boldsymbol{k})+\Delta E_{\text{Cas}}^{T=0},
\end{equation*}where the interaction term is
\begin{equation*}
\begin{split}
\Delta E_{\text{Cas}}^{T=0}=&\frac{\hbar}{2\pi}\sum_{\boldsymbol{k}\in\mathcal{S}_{\text{TM}}}\int_0^{\infty}
\log\Biggl\{\left(1-\Delta_{12}^{\text{TM}}(i\xi,\boldsymbol{k})e^{-2q_1(\xi,\boldsymbol{k})d_1}\right)
\left(1-\Delta_{32}^{\text{TM}}(i\xi,\boldsymbol{k})e^{-2q_3(\xi,\boldsymbol{k})d_3}\right) \\&-\left(  e^{-2q_1(\xi,\boldsymbol{k})d_1}-\Delta_{12}^{\text{TM}}(i\xi,\boldsymbol{k})\right)\left( e^{-2q_3(\xi,\boldsymbol{k})d_3} -\Delta_{32}^{\text{TM}}(i\xi,\boldsymbol{k})\right)e^{-2q_2(\xi,\boldsymbol{k})d_2}\Biggr\}d\xi +\left( \text{TM}\longrightarrow \text{TE}\right).
\end{split}
\end{equation*}

 In the    limit where the two ends of the rectangular box $x_1=0$ and $x_1=L$ are brought to infinity, i.e., $d_1\rightarrow \infty$ and $d_3\rightarrow \infty$, the interaction term of the Casimir energy  \eqref{eq8_3_12}  gives
\begin{equation*}
\begin{split}
\Delta E_{\text{Cas}}=& \frac{k_BT}{2}\sum_{\boldsymbol{k}\in\mathcal{S}_{\text{TM}}}\sum_{l=-\infty}^{\infty}\log \left(1 -  \Delta_{12}^{\text{TM}}(i\xi_l,\boldsymbol{k}) \Delta_{32}^{\text{TM}}(i\xi_l,\boldsymbol{k}) e^{-2q_2(\xi_l,\boldsymbol{k})d_2}\right) +\left( \text{TM}\longrightarrow \text{TE}\right).
\end{split}
\end{equation*}In addition, if  the media are infinite in the transversal direction, i.e., $L_2,L_3\rightarrow \infty$, then  $\lambda_{\boldsymbol{k}}^2$ has to be replaced by $k^2$ and  the summation over $\boldsymbol{k}$ has to be changed to an appropriate integral. More precisely, in the limit $L_2,L_3\rightarrow \infty$, the interaction term of the Casimir energy density is
\begin{equation}\label{eq8_11_8}
\begin{split}
\Delta \mathcal{E}_{\text{Cas}}= \frac{k_BT}{4\pi}\sum_{l=-\infty}^{\infty}\int_0^{\infty} \left\{\log \left(1 -  \Delta_{12}^{\text{TM}}(i\xi_l,k) \Delta_{32}^{\text{TM}}(i\xi_l,k) e^{-2q_2(\xi_l,k)d_2}\right)+\left(\text{TM}\longrightarrow \text{TE}\right)\right\}kdk,
\end{split}
\end{equation}where $q_j(\xi, k)=\sqrt{\varepsilon_j(i\xi)\mu_j(i\xi)\xi^2+k^2}$.
This is precisely the Lifshitz formula \cite{3} for the Casimir energy density when two semi-infinite media with permittivities and permeabilities $\vep_1,\mu_1$ and $\vep_3,\mu_3$ are separated by a medium of permittivity $\varepsilon_2$ and permeability $\mu_2$. Notice that we have   ignored (or subtracted away) the divergent part of the Casimir energy $d_1\Xi_1(t_c)+d_2\Xi_2(t_c)+d_3\Xi_3(t_c)$ and only take the limit $d_1, d_3\rightarrow \infty$ on the interaction term. Later on, we are going to show some situations where the divergences would naturally disappear when we consider the Casimir force.

\section{Casimir force acting on the piston}\label{s3}

We assume that the piston   is allowed to move freely inside the rectangular box, but it is non-deformable, i.e., its thickness $d_2=b-a$ is fixed. The position of the piston can be described by the variable $a=d_1$. As a function of $a$, $d_1=a$, $d_2$ is fixed and $d_3=L-a-d_2$. The Casimir force acting on the piston is given by
\begin{equation*}
\begin{split}
F_{\text{Cas}}^{\text{piston}}(a) = -\frac{\pa E_{\text{Cas}}}{\pa a}=-\frac{\pa E_{\text{Cas}}}{\pa d_1}+\frac{\pa E_{\text{Cas}}}{\pa d_3}.
\end{split}
\end{equation*}From \eqref{eq8_4_5}, \eqref{eq8_3_13} and  \eqref{eq8_3_12}, we find that
\begin{equation*}
F_{\text{Cas}}^{\text{piston}}(a)= \Xi_3(t_c)-\Xi_1(t_c) +\Delta F_{\text{Cas}}^{\text{piston}}(a),
\end{equation*}where $\Xi_1(t_c)$ and $\Xi_3(t_c)$ are defined by \eqref{eq8_3_13}, and the interaction term is
\begin{equation}\label{eq8_26_3}
\begin{split}
\Delta F_{\text{Cas}}^{\text{piston}}(a)=\Delta F_{\text{Cas};\text{TM}}^{  L}+\Delta F_{\text{Cas};\text{TE}}^{ L}-\Delta F_{\text{Cas};\text{TM}}^{ R}-\Delta F_{\text{Cas};\text{TE}}^{ R},
\end{split}
\end{equation}with
\begin{equation*}
\begin{split}
&\Delta F_{\text{Cas};\text{TM}}^{  L}=-k_BT\sum_{\boldsymbol{k}\in\mathcal{S}_{\text{TM}}}\sum_{l=-\infty}^{\infty} \frac{q_1e^{-2q_1 d_1} \Bigl(\Delta_{12}^{\text{TM}}
\left(1-\Delta_{32}^{\text{TM}} e^{-2q_3 d_3}\right)-\left( \Delta_{32}^{\text{TM}} -e^{-2q_3 d_3} \right)e^{-2q_2d_2}\Bigr)}{ \left(1-\Delta_{12}^{\text{TM}}
e^{-2q_1 d_1}\right) \left(1-\Delta_{32}^{\text{TM}} e^{-2q_3 d_3}\right)
  -\left( \Delta_{12}^{\text{TM}}  -e^{-2q_1 d_1}\right)\left( \Delta_{32}^{\text{TM}} -e^{-2q_3 d_3} \right)e^{-2q_2 d_2}},
\end{split}
\end{equation*} $\Delta F_{\text{Cas};\text{TM}}^{  R}$ is obtained from $\Delta F_{\text{Cas};\text{TM}}^{  L}$ by interchanging the index $1$ with $3$, and  $\Delta F_{\text{Cas};\text{TE}}^{  L/R}$ is obtained from $\Delta F_{\text{Cas};\text{TM}}^{  L/R}$ by changing $\Delta_{jk}^{\text{TM}}$ to $\Delta_{jk}^{\text{TE}}$. Unlike the piston cases that have been considered so far, there are still divergences $\Xi_3(t_c)-\Xi_1(t_c) $ in the Casimir force coming from the self energy terms. This term will vanish if and only if $\Xi_1(t_c)=\Xi_3(t_c)$, which would happen for example when the two media separated by the piston is isorefractive.  In particular, if the two media 1 and 3 are both vacuum,    which is the case in the previous works that have  considered the piston scenarios, then there are no divergences, and regularization is not required for the Casimir force.

For the particular case where  $\vep_1=\vep_3$, $\mu=\mu_3$, the self-energy terms   $\Xi_1(t_c)$ and $\Xi_3(t_c) $ cancel and we can set the cut-off parameter $t_c$ to zero. Let $q=q_1=q_3$ and $\Delta^{\text{TM}/\text{TE}}=\Delta_{12}^{\text{TM}/\text{TE}}=\Delta_{32}^{\text{TM}/\text{TE}}$. Then     the Casimir force acting on the piston is given by \begin{equation}\label{eq8_24_1}
\begin{split}
  F_{\text{Cas}}^{  \text{piston}}=&-k_BT\sum_{\boldsymbol{k}\in\mathcal{S}_{\text{TM}}}\sum_{l=-\infty}^{\infty} \frac{q\Delta^{\text{TM}}\left(e^{-2qd_1}-e^{-2qd_3}\right) \left(1-e^{-2q_2d_2}\right)}{ \left(1-\Delta^{\text{TM}}e^{-2q  d_1}
\right) \left(1-\Delta^{\text{TM}} e^{-2q  d_3}\right)
  -\left( \Delta^{\text{TM}} - e^{-2q  d_1}\right)\left( \Delta^{\text{TM}} -e^{-2q  d_3} \right)e^{-2q_2 d_2}}\\&+\left(\text{TM}\longrightarrow \text{TE}\right).
\end{split}
\end{equation} \begin{figure}
\epsfxsize=0.49\linewidth \epsffile{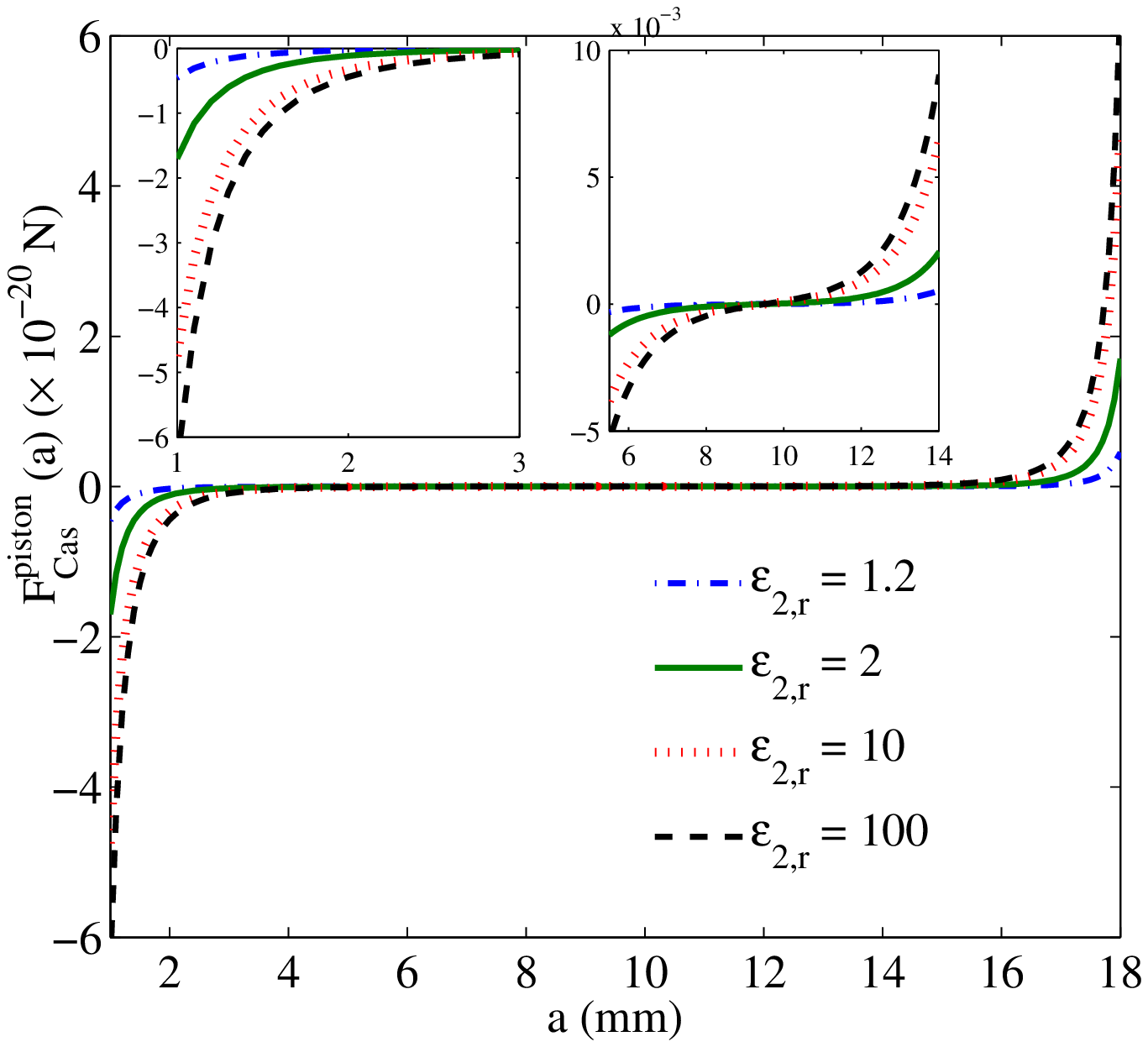} \epsfxsize=0.49\linewidth \epsffile{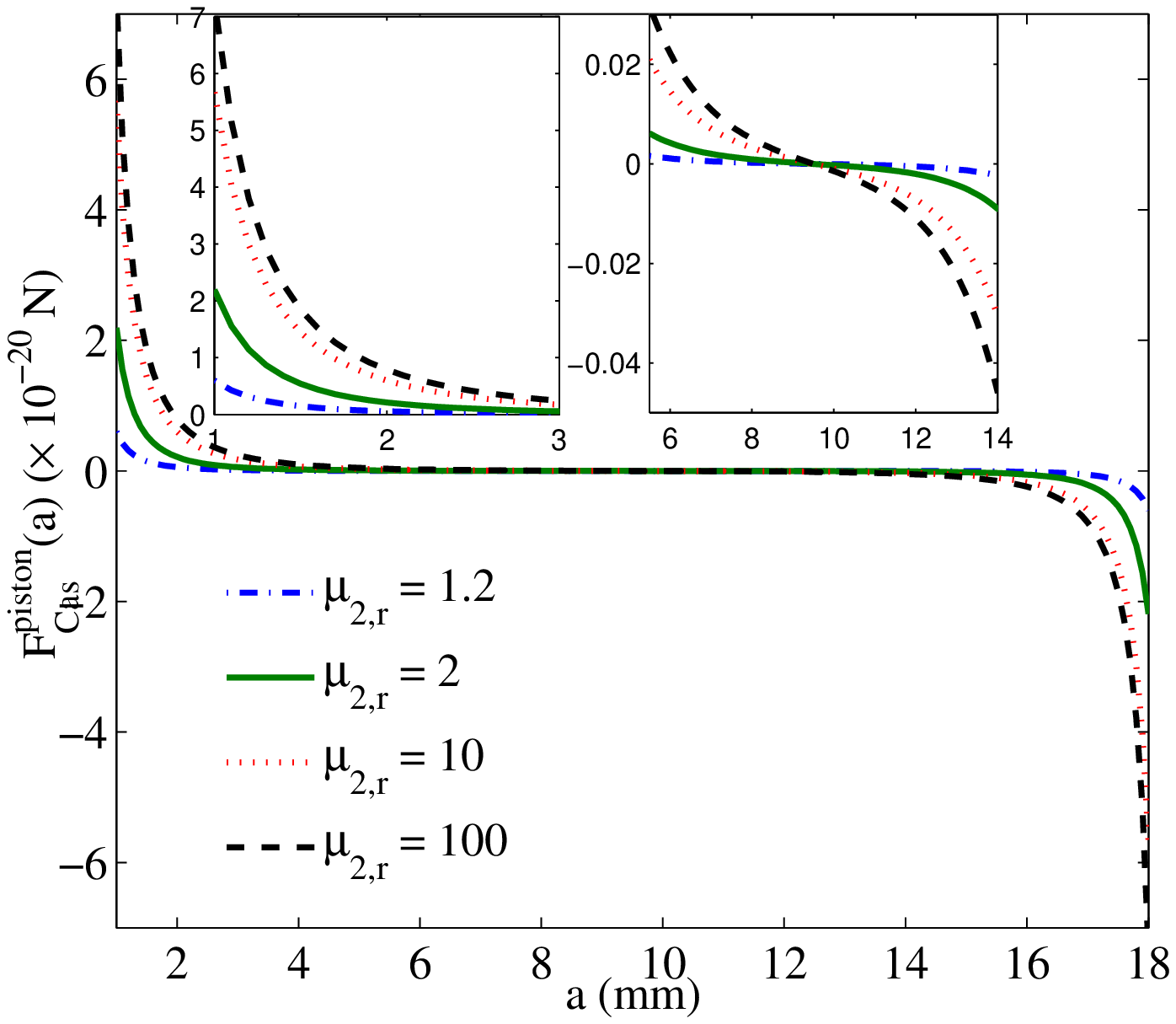} \caption{\label{f3}The Casimir force $F_{\text{Cas}}^{\text{piston}}(a)$ as a function of the plate separation $a$ for different values of $\vep_{2,r}=\vep_2/\vep_0$ and $\mu_{2,r}=\mu_2/\mu_0$.   Here   $L=2$cm, $L_2=L_3=1$cm, $d_2=1$mm and $T=1$K. The two regions separated by the piston are vacuum. For the graph on the left, $\mu_{2}=\mu_0$. For the graph on the right, $\vep_2=\vep_0$. }\end{figure}
\begin{figure}
\epsfxsize=0.49\linewidth \epsffile{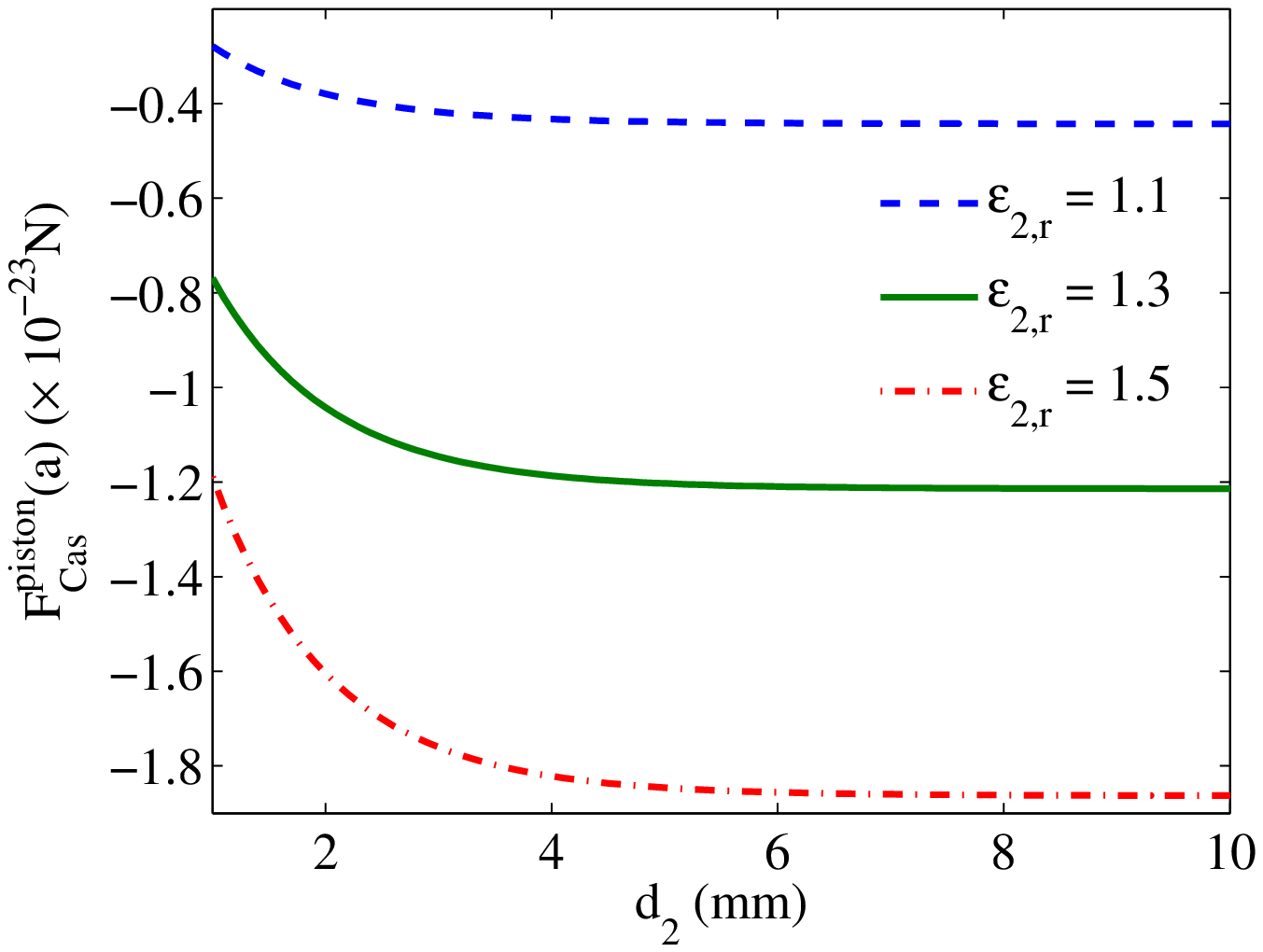} \epsfxsize=0.49\linewidth \epsffile{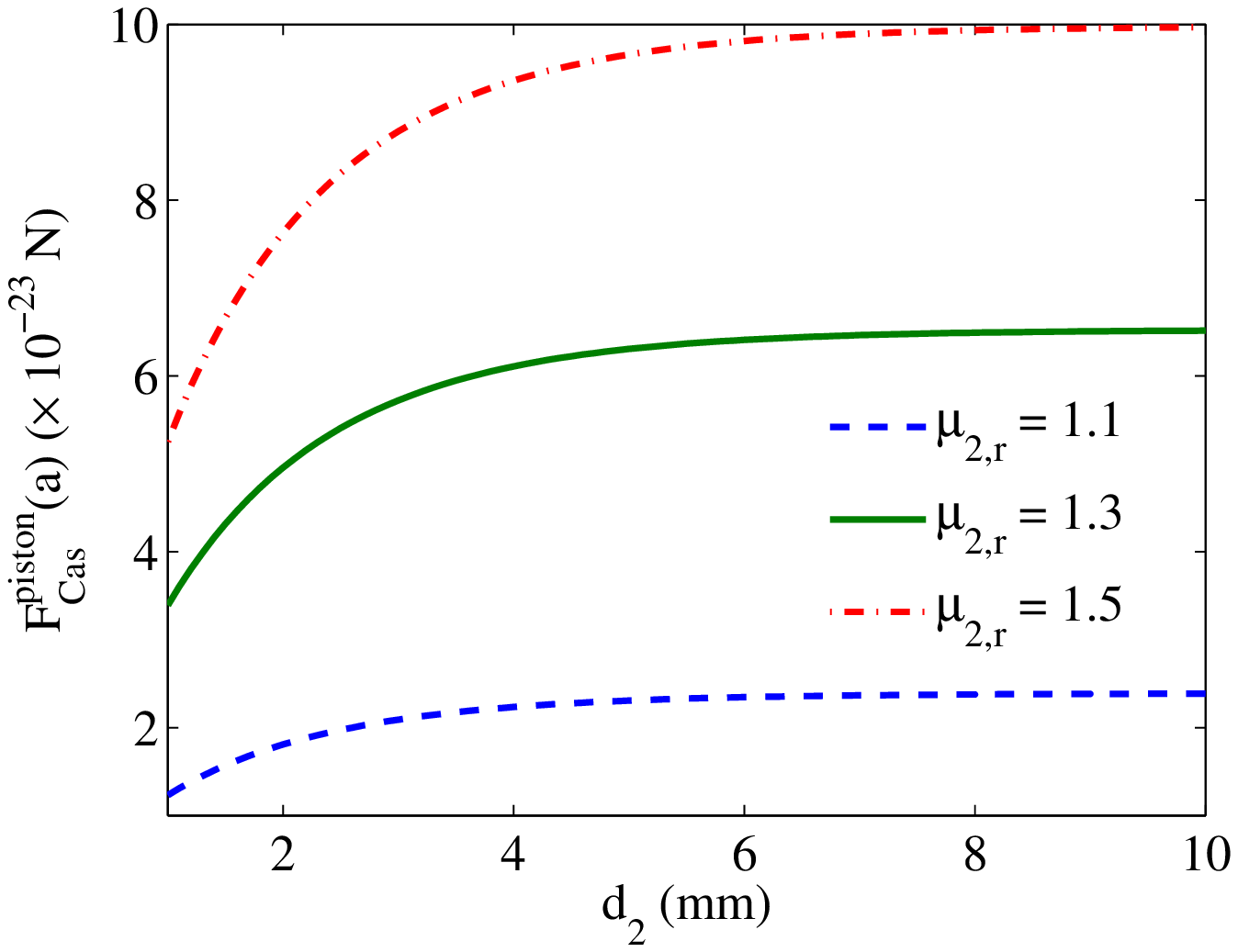} \caption{\label{f4}The Casimir force $F_{\text{Cas}}^{\text{piston}}(a)$ as a function of the piston thickness $d_2$ for different values of $\vep_{2,r}=\vep_2/\vep_0$ and $\mu_{2,r}=\mu_2/\mu_0$.   Here   $L=5$cm, $L_2=L_3=1$cm, $d_1=5$mm, $T=1$K. The two regions separated by the piston are vacuum. For the graph on the left, $\mu_{2}=\mu_0$. For the graph on the right, $\vep_2=\vep_0$. }\end{figure}From this, we find that  the sign of the Casimir force is determined by the sign of $\Delta^{\text{TM}}, \Delta^{\text{TE}}$, and the sign of $d_1-d_3$. Notice that in general the Casimir force depends on the thickness of the piston, and it vanishes as the thickness of the piston goes to zero, unless $\Delta^{\text{TM}}=\pm 1$ and $\Delta^{\text{TE}}=\pm1$. FIG. \ref{f3} and FIG. \ref{f4} show the special case where the regions separated by the piston are vacuum, i.e., $\vep_1=\vep_3=\vep_0$, $\mu_1= \mu_3=\mu_0$ and the piston is made of  material with constant permittivity $\vep_2$ and permeability $\mu_2$.  We find that when $\vep_2>\vep_0$, the Casimir force attracts the piston to the closer wall, and when $\mu_2>\mu_0$, the Casimir force push the piston away towards the middle. This can also be directly verified from \eqref{eq8_24_1}.

In the special case where the regions separated by the piston are vacuum, i.e., $\vep_1=\vep_3=\vep_0$, $\mu_1= \mu_3=\mu_0$,  and the piston is perfectly conducting,  achieved by the limit $\vep_2\rightarrow \infty$, we find that
\begin{equation}\label{eq8_4_6}\begin{split}
\Delta_{12}^{\text{TM}}(i\xi_l, \boldsymbol{k})=\Delta_{32}^{\text{TM}}(i\xi_l, \boldsymbol{k})=\frac{\vep_2\sqrt{\frac{\xi_l^2}{c^2}+\lambda_{\boldsymbol{k}}^2}-\vep_0 \sqrt{\vep_2\mu_2\xi_l^2+\lambda_{\boldsymbol{k}}^2}}{\vep_2\sqrt{\frac{\xi_l^2}{c^2}+\lambda_{\boldsymbol{k}}^2}+\vep_0 \sqrt{\vep_2\mu_2\xi_l^2+\lambda_{\boldsymbol{k}}^2}}\longrightarrow & 1,\\\Delta_{12}^{\text{TE}}(i\xi_l, \boldsymbol{k})=\Delta_{32}^{\text{TE}}(i\xi_l, \boldsymbol{k})=\frac{\mu_0 \sqrt{\vep_2\mu_2\xi_l^2+\lambda_{\boldsymbol{k}}^2}-\mu_2\sqrt{\frac{\xi_l^2}{c^2}+\lambda_{\boldsymbol{k}}^2}}{\mu_0 \sqrt{\vep_2\mu_2\xi_l^2+\lambda_{\boldsymbol{k}}^2}+\mu_2\sqrt{\frac{\xi_l^2}{c^2}+\lambda_{\boldsymbol{k}}^2}}\longrightarrow &1.
\end{split}\end{equation}When $l=0$,   the limit  of the  second term has to be taken using the prescription of Schwinger, DeRaad and Milton \cite{4} where  one first takes the limit $\vep_2\rightarrow \infty$ before setting $l=0$. Alternatively, one can also take the limit $\mu_2\rightarrow 0$. In fact,   superconductors can be considered as perfectly conducting and it is well-known that for superconductors, the magnetic permeability is zero.  Substituting the limits \eqref{eq8_4_6} into the Casimir force \eqref{eq8_24_1}, we find that
\begin{equation}\label{eq8_4_7}
F_{\text{Cas}}^{  \text{piston}}=-k_BT\left(\sum_{\boldsymbol{k}\in\mathcal{S}_{\text{TM}}}+\sum_{\boldsymbol{k}\in\mathcal{S}_{\text{TE}}}\right)\sum_{l=-\infty}^{\infty} \left\{\frac{\sqrt{\left(\frac{2\pi l k_BT}{\hbar c}\right)^2+\lambda_{\boldsymbol{k}}^2}}{ \exp\left(2d_1\sqrt{\left(\frac{2\pi l k_BT}{\hbar c}\right)^2+\lambda_{\boldsymbol{k}}^2}\right)-1}-\frac{\sqrt{\left(\frac{2\pi l k_BT}{\hbar c}\right)^2+\lambda_{\boldsymbol{k}}^2}}{ \exp\left(2d_3\sqrt{\left(\frac{2\pi l k_BT}{\hbar c}\right)^2+\lambda_{\boldsymbol{k}}^2}\right)-1} \right\}.
\end{equation}The two terms in the brackets represent the Casimir energies from the left region (with thickness $d_1=a$) and from the right region (with thickness $d_2=L-a-d_2$) respectively. This formula is exactly the same as the formula  for a perfectly conducting piston moving freely inside a perfectly conducting rectangular cavity derived in \cite{7, 6}. Notice that this formula is independent of the thickness of the piston $d_2$.

In the case the piston is infinitely permeable,  the limit is obtained by $\mu_2\rightarrow\infty$. As dual to the perfectly conducting case, we stipulate that $\vep_2\rightarrow 0$. Then
\begin{equation*}
\Delta_{12}^{\text{TM}}(i\xi_l, \boldsymbol{k})\longrightarrow -1, \hspace{1cm}\Delta_{12}^{\text{TE}}(i\xi_l, \boldsymbol{k})\longrightarrow -1.
\end{equation*} In this limit,   the Casimir force \eqref{eq8_24_1} becomes
\begin{equation}\label{eq8_4_8}
F_{\text{Cas}}^{  \text{piston}}=k_BT\left(\sum_{\boldsymbol{k}\in\mathcal{S}_{\text{TM}}}+\sum_{\boldsymbol{k}\in\mathcal{S}_{\text{TE}}}\right)\sum_{l=-\infty}^{\infty} \left\{\frac{\sqrt{\left(\frac{2\pi l k_BT}{\hbar c}\right)^2+\lambda_{\boldsymbol{k}}^2}}{ \exp\left(2d_1\sqrt{\left(\frac{2\pi l k_BT}{\hbar c}\right)^2+\lambda_{\boldsymbol{k}}^2}\right)+1}-\frac{\sqrt{\left(\frac{2\pi l k_BT}{\hbar c}\right)^2+\lambda_{\boldsymbol{k}}^2}}{ \exp\left(2d_3\sqrt{\left(\frac{2\pi l k_BT}{\hbar c}\right)^2+\lambda_{\boldsymbol{k}}^2}\right)+1} \right\},
\end{equation}which coincides with the result for an infinitely permeable piston moving freely inside a perfectly conducting piston we derived in \cite{8}. Notice again that the Casimir force is independent of the thickness of the piston $d_2$.

As one can see, in general the nature and the strength of the interaction term of the Casimir force acting on the piston \eqref{eq8_26_3} depend on the properties of the materials and the thickness of each material. It does not decouple into the Casimir force acting from the left region and the Casimir force acting from the right region.
In the limit the right end of the rectangular box is moved to infinity, i.e., $d_3\rightarrow \infty$,  we find that $\Delta F_{\text{Cas};\text{TM}}^{ R}\rightarrow 0, \Delta F_{\text{Cas};\text{TE}}^{ R}\rightarrow 0$, and
\begin{equation}\label{eq8_5_1}\begin{split}
\Delta F_{\text{Cas};\text{TM}}^{  \text{piston}, d_3\rightarrow\infty}=-k_BT\sum_{\boldsymbol{k}\in\mathcal{S}_{\text{TM}}}\sum_{l=-\infty}^{\infty} q_1
 \left(\frac{\left(1-
\Delta_{12}^{\text{TM}}  \Delta_{32}^{\text{TM}}  e^{-2q_2d_2}\right)}{\left(\Delta_{12}^{\text{TM}} -\Delta_{32}^{\text{TM}} e^{-2q_2 d_2}\right)}e^{2q_1 d_1}-1\right)^{-1} +\left(\text{TM}\longrightarrow \text{TE}\right).
\end{split}\end{equation}This formula can be interpreted as the interaction term of the Casimir force acting on a perfectly conducting plate and   a plate made  of real material (with permittivity $\vep_2$ and permeability $\mu_2$) which are   embedded  orthogonally   in an infinitely long rectangular cylinder, with the medium between them having permittivity $\vep_1$ and permeability $ \mu_1$ and the medium outside them   having permittivity $\vep_3$ and permeability $\mu_3$. \eqref{eq8_5_1} shows that when the separation between the plates $d_1$ gets large, the magnitude of the interaction term of the Casimir force decreases exponentially.

\section{Casimir force between two perfectly conducting plates separated by medium with constant refractive index}
If  both plates are perfectly conducting so that $\Delta_{12}^{\text{TM}}=\Delta_{32}^{\text{TM}}=\Delta_{12}^{\text{TE}}=\Delta_{32}^{\text{TE}}=1$, and the medium 1 between the plates has constant  refractive index $n_1=c\sqrt{\vep_1\mu_1}$, then we find that the interaction term of the Casimir force \eqref{eq8_5_1} becomes
\begin{equation}\label{eq8_11_2}
\begin{split}
&\Delta F_{\text{Cas}} =-\frac{k_BT}{2}\sum_{(k_2,k_3)\in \Z^2\setminus\{\boldsymbol{0}\}}\sum_{l=-\infty}^{\infty} \frac{\sqrt{ n_1^2\left(\frac{2\pi lk_BT}{\hbar c}\right)^2+ \lambda_{\boldsymbol{k}}^2}}{
  \exp\left(2d_1\sqrt{ n_1^2\left(\frac{2\pi lk_BT}{\hbar c}\right)^2+ \lambda_{\boldsymbol{k}}^2}\right)-1}.
\end{split}
\end{equation}This term is always attractive, and its magnitude   gets smaller if the medium between the perfectly conducting plates has higher refractive index.
In the zero temperature limit,  the interaction term of the  Casimir force \eqref{eq8_11_2} is
\begin{equation*}
\begin{split}
\Delta F_{\text{Cas}}^{ T=0}=&-\frac{\hbar c}{n_1}\Biggl(\frac{1}{4  d_1}\sum_{k_1=1}^{\infty}  \sum_{(k_2,k_3)\in \Z^2\setminus\{\boldsymbol{0}\}}\frac{\sqrt{\left(\frac{k_2}{L_2}\right)^2+\left(\frac{k_3}{L_3}\right)^2}}{k_1}K_1\left( 2\pi k_1d_1\sqrt{\left(\frac{k_2}{L_2}\right)^2+\left(\frac{k_3}{L_3}\right)^2}\right)\\&+\frac{1}{2}\sum_{k_1=1}^{\infty}  \sum_{(k_2,k_3)\in \Z^2\setminus\{\boldsymbol{0}\}}\left(\left(\frac{k_2}{L_2}\right)^2+\left(\frac{k_3}{L_3}\right)^2\right)K_0\left( 2\pi k_1d_1\sqrt{\left(\frac{k_2}{L_2}\right)^2+\left(\frac{k_3}{L_3}\right)^2}\right)\Biggr),
\end{split}
\end{equation*}which is $n_1$ times smaller than the corresponding Casimir force on perfectly conducting plates separated by vacuum. When the plate separation $d_1$ is much smaller than the size of the cross section, i.e., $d_1\ll L_2, L_3$, the leading terms of the interaction term of the Casimir force is
\begin{equation*}\begin{split}
\Delta F_{\text{Cas}} =&-\frac{\hbar c \pi^2 L_2L_3}{240 n_1d_1^4},\hspace{2cm} \text{if}\;\; T=0,\\
\Delta F_{\text{Cas}} =&-\frac{\zeta_R(3)L_2L_3k_BT}{4\pi d_1^3},\hspace{1cm}\text{if}\;\; d_1T\gg 1.\end{split}
\end{equation*}Notice that when $T=0$, the leading term is $n_1$ times smaller than the corresponding term for perfectly conducting plates separated by vacuum, but when the temperature is large enough, the leading term becomes independent of the refractive index of the medium between the plates.

The interaction term \eqref{eq8_11_2} has not taken into full account the influence of the material between the plates. There is an additional contribution to the Casimir force arising from the difference of materials in the region between the plates and the region outside the plates, given by
\begin{equation*}
\Xi_3(t_c)-\Xi_1(t_c)=\frac{1}{\pi}\left(\sum_{\boldsymbol{k}\in\mathcal{S}_{\text{TM}}}+\sum_{\boldsymbol{k}\in\mathcal{S}_{\text{TE}}}\right)\int_{p_3(\omega,\boldsymbol{k})\geq 0}\left\{
\frac{\hbar\omega}{2} e^{-t_c\omega}+k_BT\log\left(1-\exp\left(-\frac{\hbar\omega}{k_BT}\right)\right)\right\}dp_3(\omega,\boldsymbol{k})-\left(p_3\longrightarrow p_1\right).
\end{equation*}
Assume that the permittivity $\vep_3$ and the permeability $\mu_3$ of the medium in region 3 are also constants with refractive index $n_3$. Then  up to the term constant in $t_c$,
\begin{equation}\label{eq12_07_3}\begin{split}
 \Xi_1(t_c)=&\frac{3\hbar L_2L_3}{\pi^2c^3}n_1^3t_c^{-4} -\frac{\hbar n_1 t_c^{-2}}{4\pi c} -\frac{\hbar c}{n_1}\left(\frac{\pi^2}{720}\frac{L_3}{L_2^3}+\frac{\zeta_R(3)}{16\pi L_3^2}+\frac{1}{2L_2^{\frac{3}{2}}L_3^{\frac{1}{2}}}\sum_{k_2=1}^{\infty}\sum_{k_3=1}^{\infty}\left(\frac{k_2}{k_3}\right)^{\frac{3}{2}}K_{\frac{3}{2}}\left(\frac{2\pi k_2k_3L_3}{L_2}\right)\right)\\&-\frac{k_BT}{2}\sum_{l=1}^{\infty}\sum_{(k_2,k_3)\in \Z^2\setminus\{\boldsymbol{0}\}}\frac{\sqrt{\left(\frac{k_2}{L_2}\right)^2+\left(\frac{k_3}{L_3}\right)^2}}{l}K_1\left( \frac{\pi \hbar c l}{n_1k_BT}\sqrt{\left(\frac{k_2}{L_2}\right)^2+\left(\frac{k_3}{L_3}\right)^2}\right),
\end{split}\end{equation}and $\Xi_3(t_c)$ is obtained from this formula by replacing $n_1$ with $n_3$.  It is obvious that if $n_1\neq n_3$, the terms  $\Xi_1(t_c)$ and $\Xi_3(t_c)$ due to the self energy of the system are not equal.  As explained in \cite{new2, new3, new4, new5}, in this case, the parameter $t_c$ has to be present in the expressions for Casimir energy and Casimir force.
\section{The five-layer model}
\begin{figure}
\epsfxsize=0.5\linewidth \epsffile{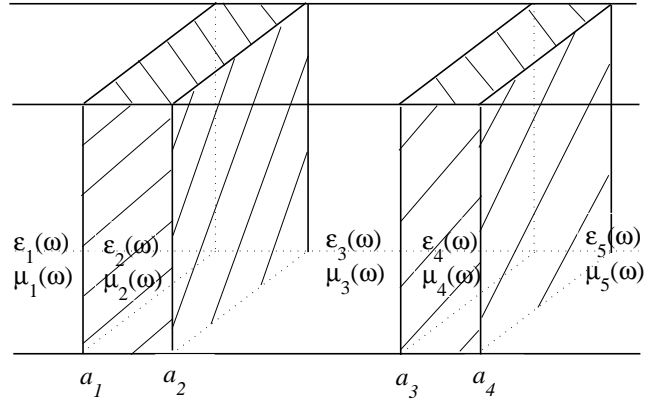} \caption{\label{f2}A five-layer model }\end{figure}
The piston system we consider in Sections \ref{s2} and \ref{s3} can be considered as a regularization device for a three-layer model. We have asserted that in the limit where one end of the rectangular box is brought to infinity,  we obtain the Casimir force acting on a pair of parallel plates, one being perfectly conducting and the other made of real material.  In this section, we consider the more general case of the five-layer model (FIG. \ref{f2}) which is the more appropriate model for two parallel plates both made of real materials.

The Casimir energy and Casimir force of a five-layer model have been derived by some authors using different methods \cite{new1, 5,9, 10, 11, 12, 13, 54}. We are going to use the piston approach. Namely, first we assume that the whole system is inside a large perfectly conducting rectangular cavity $[0,L]\times[0,L_2]\times [0,L_3]$. As depicted in FIG. \ref{f2}, there are five layers of materials with electric permittivities $\vep_1,\vep_2, \vep_3,\vep_4, \vep_5$ and magnetic permeabilities $\mu_1,\mu_2,\mu_3,\mu_4, \mu_5$ from left to right. The first medium extends from $x_1=0$ to $x_1=a_1$, the second from $x_1=a_1$ to $x_1=a_2$, the third from $x_1=a_2$ to $x_1=a_3$, the fourth from $x_1=a_3$ to $x_1=a_4$ and the last one from $x_1=a_4$ to $x_1=L$.

Let $d_1=a_1, d_2=a_2-a_1, d_3=a_3-a_2, d_4=a_4-a_3$ and $d_5=L-a_4$. Then using the same method as Section \ref{s2}, we find that the Casimir energy of the system is
\begin{equation}\label{eq8_26_4}
E_{\text{Cas}}(t_c)=\sum_{j=1}^5 d_j\Xi_j(t_c)+\Delta E_{\text{Cas}},
\end{equation}where $\Xi_j(t_c)$ is as in \eqref{eq8_3_13}, and the interaction term of the Casimir energy is
\begin{equation*}\begin{split}
&\Delta E_{\text{Cas}}=\frac{k_BT}{2}\sum_{\boldsymbol{k}\in \mathcal{S}_{\text{TM}}}\sum_{l=-\infty}^{\infty} \log\Biggl\{\Bigl(\left[1+\Delta^{\text{TM}}_{23}\Delta^{\text{TM}}_{12}e^{-2q_2d_2}\right]
-\left[\Delta^{\text{TM}}_{12}+\Delta^{\text{TM}}_{23}
e^{-2q_2d_2}\right]e^{-2q_1d_1}\Bigr)\\&\times\Bigl(\left[1+\Delta^{\text{TM}}_{43}\Delta^{\text{TM}}_{54}e^{-2q_4d_4}\right]-\left[\Delta^{\text{TM}}_{54}+\Delta^{\text{TM}}_{43}
e^{-2q_4d_4}\right]e^{-2q_5d_5}\Bigr)-\Bigl(\left[e^{-2q_2d_2}+\Delta^{\text{TM}}_{23}\Delta^{\text{TM}}_{12}\right]e^{-2q_1d_1}-\left[\Delta^{\text{TM}}_{12}e^{-2q_2d_2}
+\Delta^{\text{TM}}_{23}
\right]\Bigr)\\&\times\Bigl(\left[e^{-2q_4d_4}+\Delta^{\text{TM}}_{43}\Delta^{\text{TM}}_{54}\right]e^{-2q_5d_5}-\left[\Delta^{\text{TM}}_{54}e^{-2q_4d_4}+\Delta^{\text{TM}}_{43}
\right]\Bigr)e^{-2q_3d_3}\Biggr\} +\left(\text{TM}\longrightarrow\text{TE}\right), \\& \text{where}\;\; q_j=q_j(\xi_l,\boldsymbol{k}), \;\; \Delta^{\text{TM}}_{jk}=\Delta^{\text{TM}}_{jk}(i\xi_l,\boldsymbol{k}),\;\; \Delta^{\text{TE}}_{jk}=\Delta^{\text{TE}}_{jk}(i\xi_l,\boldsymbol{k}).
\end{split}\end{equation*}The functions $q_j(\xi_l,\boldsymbol{k})$, $\Delta^{\text{TM}}_{jk}=\Delta^{\text{TM}}_{jk}(i\xi_l,\boldsymbol{k})$ and $\Delta^{\text{TE}}_{jk}=\Delta^{\text{TE}}_{jk}(i\xi_l,\boldsymbol{k})$ are defined as in \eqref{eq8_5_2}.

 Now assume that   medium 2 and medium 4 are two non-deformable pistons/plates embedded orthogonally inside  a rectangular box, dividing it into three regions filled with different materials. The Casimir forces acting on the plate on the left (medium 2) and the plate on the right (medium 4) are given respectively by
\begin{equation*}\begin{split}
 F_{\text{Cas}}^L=&-\frac{\pa   E_{\text{Cas}}(t_c)}{\pa d_1}+\frac{\pa  E_{\text{Cas}}(t_c)}{\pa d_3}=-\Xi_1(t_c)+\Xi_3(t_c)-\frac{\pa  \Delta E_{\text{Cas}} }{\pa d_1}+\frac{\pa \Delta  E_{\text{Cas}} }{\pa d_3},\\
 F_{\text{Cas}}^R=&-\frac{\pa   E_{\text{Cas}}(t_c)}{\pa d_3}+\frac{\pa  E_{\text{Cas}}(t_c)}{\pa d_5}=-\Xi_3(t_c)+\Xi_5(t_c)-\frac{\pa  \Delta E_{\text{Cas}} }{\pa d_3}+\frac{\pa \Delta  E_{\text{Cas}} }{\pa d_5}.
\end{split}\end{equation*}
By moving the two auxiliary ends of the rectangular box to infinity, i.e., $d_1,d_5\rightarrow \infty$, since $$\frac{\pa \Delta  E_{\text{Cas}} }{\pa d_1}\xrightarrow{d_1\rightarrow \infty} 0, \;\;\frac{\pa \Delta  E_{\text{Cas}} }{\pa d_5}\xrightarrow{d_5\rightarrow \infty} 0,$$
we find that\begin{equation*}\begin{split}
 F_{\text{Cas}}^L=& -\Xi_1(t_c)+\Xi_3(t_c) +\Delta F_{\text{Cas}},\\
 F_{\text{Cas}}^R=& -\Xi_3(t_c)+\Xi_5(t_c)-\Delta F_{\text{Cas}},
\end{split}\end{equation*}  where the interaction term of the Casimir force acting on the two plates is
\begin{equation}\label{eq8_5_4}
\begin{split}
&\Delta F_{\text{Cas}}=-\frac{\pa  \Delta E_{\text{Cas}} }{\pa d_3}(d_1, d_5\rightarrow \infty)\\=&- k_BT \sum_{\boldsymbol{k}\in \mathcal{S}_{\text{TM}}}\sum_{l=-\infty}^{\infty}  q_3\left(\frac{\left( 1+\Delta^{\text{TM}}_{23}\Delta^{\text{TM}}_{12}e^{-2q_2d_2} \right)\left( 1+\Delta^{\text{TM}}_{43}\Delta^{\text{TM}}_{54}e^{-2q_4d_4} \right) }{\left(\Delta^{\text{TM}}_{12}e^{-2q_2d_2}+\Delta^{\text{TM}}_{23}
 \right)\left(  \Delta^{\text{TM}}_{54}e^{-2q_4d_4}+\Delta^{\text{TM}}_{43}
 \right)}e^{2q_3d_3} -1\right)^{-1} +\left(\text{TM}\longrightarrow \text{TE}\right).
\end{split}
\end{equation} The divergences $-\Xi_1(t_c)+\Xi_3(t_c)$ and $-\Xi_3(t_c)+\Xi_5(t_c)$   come from the self energy of the system and in general do not vanish. A special case they would vanish is that the media 1, 3 and 5 are isorefractive. In the symmetric setup, i.e.,  media 1 and 5 are isorefractive, then the force $-\Xi_1(t_c)+\Xi_3(t_c)$ acting on the plate on the left  and the force $-\Xi_3(t_c)+\Xi_5(t_c)$ acting on the plate on the right have opposite signs.
In the limit the plate on the left is perfectly conducting, i.e., $\vep_2\rightarrow \infty$, then $\Delta^{\text{TM}}_{12}, \Delta^{\text{TE}}_{12}\rightarrow 1$ and $\Delta^{\text{TM}}_{23}, \Delta^{\text{TE}}_{23}\rightarrow -1$, we find that the formula \eqref{eq8_5_4} coincides with the formula \eqref{eq8_5_1} after the re-indexing $3\rightarrow 1, 4\rightarrow 2$ and $5\rightarrow 3$. This is consistent with our interpretation of formula \eqref{eq8_5_1} as the Casimir force acting between a perfectly conducting plate and a plate made of real material.

Usually in the works of Casimir effect on parallel plates, it is assumed that the size of the cross section of the plates is much larger than the separation between the plates. In this limit, the interaction term of the Casimir force density acting on the plates is
\begin{equation}\label{eq8_26_5}
\begin{split}
\Delta \mathcal{F}_{\text{Cas}}=&- \frac{k_BT }{2\pi} \sum_{l=-\infty}^{\infty} \int_0^{\infty} q_3\left(\frac{\left( 1+\Delta^{\text{TM}}_{23}\Delta^{\text{TM}}_{12}e^{-2q_2d_2} \right)\left( 1+\Delta^{\text{TM}}_{43}\Delta^{\text{TM}}_{54}e^{-2q_4d_4} \right) }{\left(\Delta^{\text{TM}}_{12}e^{-2q_2d_2}+\Delta^{\text{TM}}_{23}
 \right)\left(  \Delta^{\text{TM}}_{54}e^{-2q_4d_4}+\Delta^{\text{TM}}_{43}
 \right)}e^{2q_3d_3} -1\right)^{-1}kdk +\left(\text{TM}\longrightarrow \text{TE}\right).
\end{split}
\end{equation}
In the zero temperature limit, \eqref{eq8_5_4} gives
\begin{equation}\label{eq8_11_10}
\begin{split}
\Delta F_{\text{Cas}}^{T=0}= &- \frac{\hbar}{\pi}\sum_{\boldsymbol{k}\in \mathcal{S}_{\text{TM}}}\int_0^{\infty}  q_3\left(\frac{\left( 1+\Delta^{\text{TM}}_{23}\Delta^{\text{TM}}_{12}e^{-2q_2d_2} \right)\left( 1+\Delta^{\text{TM}}_{43}\Delta^{\text{TM}}_{54}e^{-2q_4d_4} \right) }{\left(\Delta^{\text{TM}}_{12}e^{-2q_2d_2}+\Delta^{\text{TM}}_{23}
 \right)\left(  \Delta^{\text{TM}}_{54}e^{-2q_4d_4}+\Delta^{\text{TM}}_{43}
 \right)}e^{2q_3d_3} -1\right)^{-1}d\xi +\left(\text{TM}\longrightarrow \text{TE}\right).
\end{split}
\end{equation}In addition, if $d_3\ll L_2, L_3$, then the zero temperature Casimir force density acting on the plates is
\begin{equation}\label{eq8_11_12}
\begin{split}
\Delta \mathcal{F}_{\text{Cas}}^{T=0}= &- \frac{\hbar}{2\pi^2}\int_0^{\infty}\int_0^{\infty}  q_3\left\{\left(\frac{\left( 1+\Delta^{\text{TM}}_{23}\Delta^{\text{TM}}_{12}e^{-2q_2d_2} \right)\left( 1+\Delta^{\text{TM}}_{43}\Delta^{\text{TM}}_{54}e^{-2q_4d_4} \right) }{\left(\Delta^{\text{TM}}_{12}e^{-2q_2d_2}+\Delta^{\text{TM}}_{23}
 \right)\left(  \Delta^{\text{TM}}_{54}e^{-2q_4d_4}+\Delta^{\text{TM}}_{43}
 \right)}e^{2q_3d_3} -1\right)^{-1}+\left(\text{TM}\longrightarrow \text{TE}\right)\right\}d\xi kdk.
\end{split}
\end{equation}

Formulas \eqref{eq8_5_4}, \eqref{eq8_26_5}, \eqref{eq8_11_10} and \eqref{eq8_11_12} can be used to study the nature and properties of the Casimir force acting on two plates made of real materials. We would like to emphasize that in the case the materials in the regions 1, 3, 5 separated by the plates are not isorefractive, one   also has to take into account the forces coming from the terms $\Xi_1, \Xi_3$ and $\Xi_5$ and the cut-off parameter $t_c$ cannot be set to zero.

\begin{figure}
\epsfxsize=0.49\linewidth \epsffile{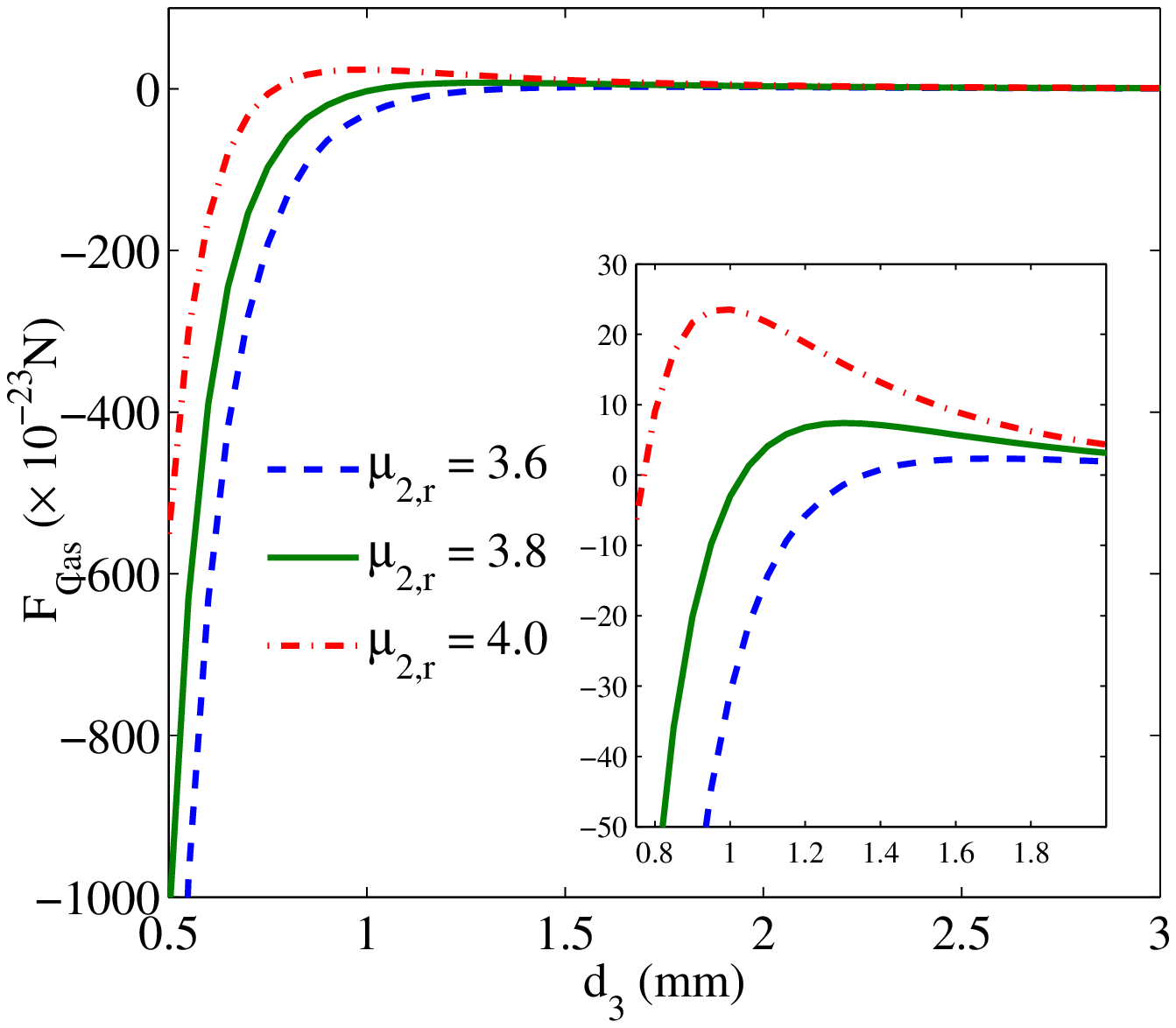} \epsfxsize=0.49\linewidth \epsffile{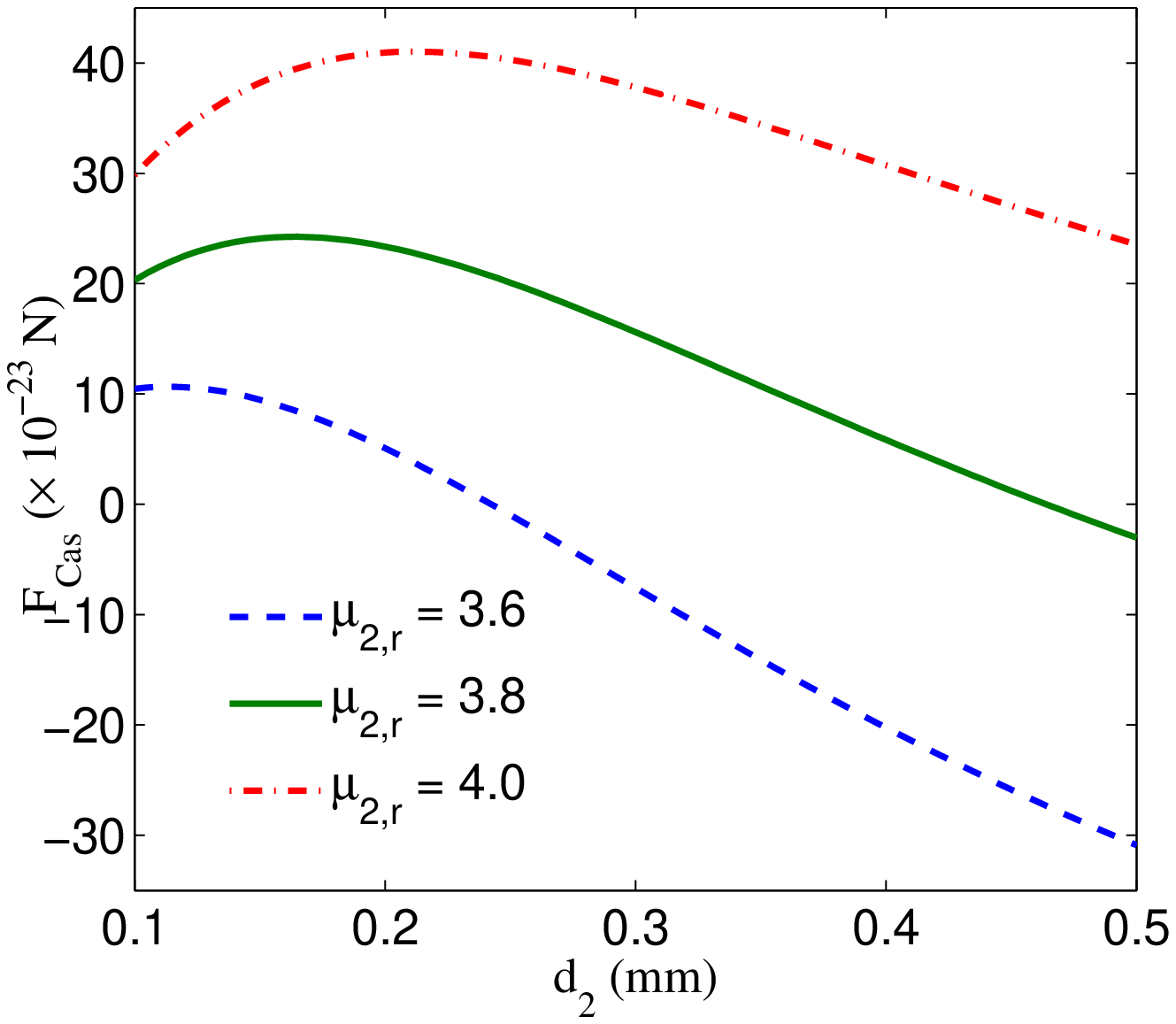} \caption{\label{f6}The Casimir force $F_{\text{Cas}}$ as a function of the plate separation $d_3$ (left) and the plate thickness $d_2=d_4$ (right). Here $L_2=L_3=1$cm, $T=0.1$K, the regions separated by the plates are vacuum. For the plate on the left $\vep_{2}=2\vep_0$ and $\mu_2=\mu_{2,r}\mu_0$, where $\mu_{2,r}=3.6,3.8,4.0$.  For the plate on the right, $\vep_4=4\vep_0$, $\mu_4=\mu_0$. For the graph on the left, the thickness of the plates are $d_2=d_4=0.5$mm. For the graph on the right, the separation between the plates is $d_3=1$mm.  }\end{figure}

In the case the three media 1, 3, 5 are the same, i.e., $\vep_1=\vep_3=\vep_5$, and $\mu_1=\mu_3=\mu_5$, we find that the self energy terms cancel and the Casimir force acting on the plates is \eqref{eq8_5_4}:
\begin{equation}\label{eq8_26_6}
\begin{split}
 F_{\text{Cas}} =&- k_BT \sum_{\boldsymbol{k}\in \mathcal{S}_{\text{TM}}}\sum_{l=-\infty}^{\infty}  q_3\left(\frac{\left( 1-\left[\Delta^{\text{TM}}_{23}\right]^2e^{-2q_2d_2} \right)\left( 1-\left[\Delta^{\text{TM}}_{43}\right]^2e^{-2q_4d_4} \right) }{\Delta^{\text{TM}}_{23}\Delta^{\text{TM}}_{43}\left(1-e^{-2q_2d_2}
 \right)\left(  1-e^{-2q_4d_4}
 \right)}e^{2q_3d_3} -1\right)^{-1} +\left(\text{TM}\longrightarrow \text{TE}\right).
\end{split}
\end{equation}In addition, if the two plates are made of the same material, i.e., $\vep_2=\vep_4$ and $\mu_2=\mu_4$, \eqref{eq8_26_6} becomes  \begin{equation}
\begin{split}
 F_{\text{Cas}} =&- k_BT \sum_{\boldsymbol{k}\in \mathcal{S}_{\text{TM}}}\sum_{l=-\infty}^{\infty}  q_3\left(\frac{\left( 1-\left[\Delta^{\text{TM}}_{23}\right]^2e^{-2q_2d_2} \right)\left( 1-\left[\Delta^{\text{TM}}_{23}\right]^2e^{-2q_2d_4} \right) }{\left[\Delta^{\text{TM}}_{23}\right]^2\left(1-e^{-2q_2d_2}
 \right)\left(  1-e^{-2q_2d_4}
 \right)}e^{2q_3d_3} -1\right)^{-1} +\left(\text{TM}\longrightarrow \text{TE}\right).
\end{split}
\end{equation}It is easy to see that in this case, the Casimir force is always attractive. This is a special case of the theorem \cite{14, 15} which asserts that the Casimir force between   two bodies with the same property is attractive. In general \eqref{eq8_26_6} shows that the dependence of the Casimir force on the thickness of the plates and the distance between the plates can be quite complicated. In FIG. \ref{f6}, we plot the dependence of the Casimir force on the distance between the plates and the thickness of the plates. The graphs show that with $\vep_2=2\vep_0$, $\vep_4=4\vep_0$, $\mu_4=\mu_0$ and $\mu_2=3.6\mu_0, 3.8\mu_0, 4.0\mu_0$,  the Casimir force can be attractive or repulsive depending on the distance between the plates and thickness of the plates.

Finally we see how we can recover the Lifshitz formula from \eqref{eq8_26_5}.
As mentioned before, when the regions separated by the two plates are filled with isorefractive media, then the divergences of the Casimir force cancel. If in addition, we assume that both plates have infinite thickness,  i.e., $d_2, d_4\rightarrow \infty$, \eqref{eq8_26_5} gives
\begin{equation*}
\begin{split}
F_{\text{Cas}}=&- \frac{k_BT }{2\pi} \sum_{l=-\infty}^{\infty} \int_0^{\infty} q_3(\xi_l, k)\left( \left[\Delta^{\text{TM}}_{23}(i\xi_l, k)\right]^{-1}
 \left[\Delta^{\text{TM}}_{43}(i\xi_l, k)\right]^{-1}
  e^{2q_3(\xi_l,k)d_3} -1\right)^{-1} kdk+\left(\text{TM}\longrightarrow \text{TE}\right).
\end{split}
\end{equation*}
This is precisely the Lifshitz formula for the Casimir force acting on two parallel plates made of real materials. Thus we see that the piston approach can be used to derive the Lifshitz formula. An advantage of this approach is that the divergence of the Casimir force is not ignored by ad hoc subtraction, but is seen to vanish due to pairwise cancelation.

\section{Conclusion} In this article, we derive the Casimir energy and Casimir force acting on a piston made of real material that moves freely inside a perfectly conducting rectangular box. We also  use the piston approach to derive the Casimir energy and Casimir force acting on two parallel plates made of real materials. We use an exponential cut-off to regularize the Casimir energy. When the regions separated by the piston or the plates are filled with isorefractive media, we show that the divergences of the Casimir force cancel each other and therefore the cut-off parameter can be set to zero. In general, the Casimir energy contains divergent terms that depend on the cut-off parameter that cannot be ignored. Although we have restricted our consideration to three-layer model (for the piston) and five-layer model (for the parallel plates), our method can be easily   generalized to Casimir effect on multi-layer models. Moreover, it is easy to see that there is nothing special in choosing the rectangular cavity as a space cut-off for the electromagnetic field. We can also use a cylinder with arbitrary cross section as a substitute. For doing so, one just has to replace  the $\lambda_{\boldsymbol{k}}^2$ by an appropriate spectrum.

For simplicity, we have illustrated the behavior of the Casimir force using non-realistic examples where the electric permittivities and magnetic permeabilities of the materials are assumed to be constants. In reality, the electric permittivities and magnetic permeabilities depend on temperature and also the frequency of the field. There are some models that have been used for the electric permittivities such as the plasma model and the Drude model. An issue that we do not discuss in this article is the leading term of the thermal correction to the Casimir force at low and high temperatures. This has been an active topic of research and the results often depend on the particular model chosen for the electric permittivities and magnetic permeabilities. In the case where the self energy terms cancel, we have shown that the high temperature leading term of the Casimir force is linear in temperature.  In case the self energy terms do not cancel, further regularization may be needed to cancel the high temperature leading terms in the self energy that are higher than linear orders \cite{new6}. This issue will be discussed in more detail elsewhere.

\begin{acknowledgments}
This project is   funded by Ministry of Science, Technology and Innovation, Malaysia under e-Science fund 06-02-01-SF0080. We would like to thank the anonymous referee for the helpful comments.
\end{acknowledgments}

\appendix
\section{Generalized Abel-Plana Summation Formula}\label{a1}

 The generalized Abel-Plana summation formula states that
if $f_0(z), f_1(z)$ and $f_2(z) $ are meromorphic functions, and
\begin{equation}\label{eq6_24_2}\begin{split}
\lim_{Y\rightarrow \infty} \int_{\alpha}^{\beta} \Bigl\{ f_0(x+iY)- f_j(x+ iY)\Bigr\}dx=0,\hspace{1cm}\text{for}\;\;j=1,2,
\end{split}
\end{equation} then by residue theorem,
\begin{equation}\label{eq6_24_1}
\begin{split}
&\sum_{\alpha\leq \text{Re}\; z\leq \beta} w_0(z) \text{Res}_{z}f_0(z) -\sum_{\substack{\alpha\leq \text{Re}\; z\leq \beta\\ \text{Im}\;z\geq 0}} w_1(z) \text{Res}_{z}f_1(z)
 -\sum_{\substack{\alpha\leq \text{Re}\; z\leq \beta\\ \text{Im}\;z\leq 0}} w_2(z) \text{Res}_{z}f_2(z) \\=&
 \frac{1}{2\pi i}\int_{L_0^++L_1^++L_2^+}(f_0(z)-f_1(z))dz+\frac{1}{2\pi i}\int_{L_0^-+L_1^-+L_2^-}(f_0(z)-f_2(z))dz\\=&  \frac{1}{2\pi }\int_0^{\infty} \Bigl\{
\left. f_0(x+iy)-f_1(x+iy)\Bigr\}\right|_{x=\alpha}^{x=\beta} dy+ \frac{1}{2\pi }\int_0^{\infty} \Bigl\{
\left. f_0(x-iy)-f_2(x-iy)\Bigr\}\right|_{x=\alpha}^{x=\beta}dy\\&-\frac{1}{2\pi i}\lim_{\epsilon\rightarrow 0^+}\int_{\alpha}^{\beta} \Bigl\{f_1(x+i\epsilon)-f_2(x-i\epsilon)\Bigr\}dx.
\end{split}
\end{equation}\begin{figure}
\epsfxsize=0.5\linewidth \epsffile{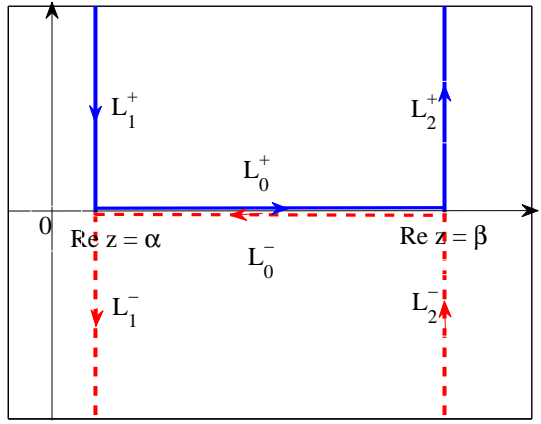}  \caption{\label{f7} The integration contour for the integrals in \eqref{eq6_24_1}.}\end{figure}\begin{equation*}
\end{equation*}Here\begin{equation*}\begin{split}w_0(z)&=\begin{cases}1, \hspace{0.5cm}&\text{if}\,\,z\in \mathfrak{D}_0, \\
 1/2, &\text{if}\,\,z\in\partial \mathfrak{D}_0,\end{cases},\hspace{0.5cm}  w_1(z) =\begin{cases} 1, \hspace{0.5cm}&\text{if}\,\,z\in \mathfrak{D}_1,\\
   1/2, &\text{ if} \,\,z\in \partial \mathfrak{D}_1\setminus\{\alpha,\beta\},\\
   1/4, &\text{if}\,\,z=\alpha\,\text{or}\,\beta,\end{cases}\hspace{0.5cm} w_2(z)=\begin{cases}1, \hspace{0.5cm}&\text{if}\,\,z\in\mathfrak{D}_2,\\
     1/2, &\text{if}\,\,z\in\partial\mathfrak{D}_2\setminus\{\alpha,\beta\},\\
   1/4, &\text{if}\,\,z=\alpha\,\text{or}\,\beta,\end{cases} \end{split}\end{equation*}
where $\mathfrak{D}_0=\left\{ z\;:\;\alpha<\text{Re}\,z<\beta\right\},\; \mathfrak{D}_1=\left\{ z\;:\;\alpha<\text{Re}\,z<\beta,\,\text{Im}z>0\right\},\; \mathfrak{D}_2=\left\{ z\;:\;\alpha<\text{Re}\,z<\beta,\,\text{Im}z<0\right\}.$ The integration contours $L_0^{\pm}, L_1^{\pm}, L_2^{\pm}$ are shown in FIG. \ref{f7}. This formula  can be proved in the same way as in \cite{1,2}.

\section{Computation of the Casimir energy}\label{a2}
In this section, we compute the Casimir energy \eqref{eq8_3_4} inside the piston system using the generalized Abel-Plana summation formula \eqref{eq6_24_1}. First we consider the zero temperature part
\begin{equation*}
E_{\text{Cas}}^{T=0}(t_c)=\frac{\hbar}{2}\sum_{\text{modes}}\omega e^{-t_c\omega}=\frac{\hbar}{2}\sum_{\text{\text{TM} modes}}\omega e^{-t_c\omega}+\frac{\hbar}{2}\sum_{\text{\text{TE} modes}}\omega e^{-t_c\omega}.
\end{equation*}Recall that the TM  modes are the positive zeros of the functions $F_{0;\text{TM}}(\omega,\boldsymbol{k}),\boldsymbol{k}=(k_2,k_3)\in \mathcal{S}_{\text{TM}}$ \eqref{eq8_3_9}. Therefore we take $\alpha=0$ and let $\beta\rightarrow\infty$ in \eqref{eq6_24_1}. Let \begin{equation*}
F_{1;\text{TM}}(\omega,\boldsymbol{k})=-r_{12}^+r_{32}^+e^{-ip_1d_1-ip_3d_3-ip_2d_2}, \hspace{1cm}F_{2;\text{TM}}(\omega,\boldsymbol{k})=r_{12}^+r_{32}^+e^{ip_1d_1+ip_3d_3+ip_2d_2},
\end{equation*}
and define
\begin{equation*}
f_{j;\text{TM}}(\omega,\boldsymbol{k})=\frac{\hbar\omega }{2}e^{-t_c\omega}\frac{d}{d\omega}\log F_{j;\text{TM}}(\omega,\boldsymbol{k}),\hspace{1cm}j=0,1,2.
\end{equation*}One can then check that the condition \eqref{eq6_24_2} is satisfied. With the further assumption that $r_{12}^+(\omega, \boldsymbol{k})$ and $r_{32}^+(\omega, \boldsymbol{k})$ do not have zeros, we find that $f_{1;\text{TM}}(\omega,\boldsymbol{k})$ and $f_{2;\text{TM}}(\omega,\boldsymbol{k})$ do not have poles. Therefore the generalized Abel-Plana summation formula \eqref{eq6_24_1} gives \begin{equation*}\label{eq12_07_1}\begin{split}& \frac{\hbar}{2}\sum_{\text{\text{TM} modes}}\omega e^{-t_c\omega} = -\frac{1}{2\pi}\sum_{\boldsymbol{k}\in\mathcal{S}_{\text{TM}}} \int_0^{\infty}\left(f_0(i\xi, \boldsymbol{k})-f_1(i\xi, \boldsymbol{k})\right)d\xi
-\frac{1}{2\pi}\sum_{\boldsymbol{k}\in\mathcal{S}_{\text{TM}}} \int_0^{\infty}\left(f_0(-i\xi, \boldsymbol{k})-f_2(-i\xi, \boldsymbol{k})\right)d\xi\\&-\frac{1}{2\pi i}\sum_{\boldsymbol{k}\in\mathcal{S}_{\text{TM}}} \lim_{\epsilon\rightarrow 0^+}\int_0^{\infty}\left(f_1(\omega+i\epsilon, \boldsymbol{k})-f_2(\omega-i\epsilon,\boldsymbol{k})\right)d\omega
\end{split}\end{equation*}For the right hand side, a straightforward computation gives
\begin{equation*}
\begin{split}
 f_0(i\xi, \boldsymbol{k})-f_1(i\xi,\boldsymbol{k})=&\frac{\hbar  \xi}{2}e^{-it_c\xi} \frac{d}{d\xi}\log\Biggl\{\left( e^{2ip_1(i\xi, \boldsymbol{k})d_1}-\frac{r_{12}^-(i\xi, \boldsymbol{k})}{r_{12}^+(i\xi, \boldsymbol{k})} \right)\left( e^{2ip_3(i\xi, \boldsymbol{k})d_3}-\frac{r_{32}^-(i\xi, \boldsymbol{k})}{r_{32}^+(i\xi, \boldsymbol{k})} \right)e^{2ip_2(i\xi, \boldsymbol{k})d_2}\\&\hspace{2cm}-\left( 1-\frac{r_{12}^-(i\xi, \boldsymbol{k})}{r_{12}^+(i\xi, \boldsymbol{k})}e^{2ip_1(i\xi, \boldsymbol{k})d_1}\right)\left(1-\frac{r_{32}^-(i\xi, \boldsymbol{k})}{r_{32}^+(i\xi, \boldsymbol{k})}e^{2ip_3(i\xi, \boldsymbol{k})d_3}\right) \Biggr\}\\
=&\frac{\hbar  \xi}{2}e^{-it_c\xi}\frac{d}{d\xi}\log\Biggl\{\left(1-\Delta_{12}^{\text{TM}}(i\xi,\boldsymbol{k})e^{-2q_1(\xi,\boldsymbol{k})d_1}\right)
\left(1- \Delta_{32}^{\text{TM}}(i\xi,\boldsymbol{k})e^{-2q_3(\xi,\boldsymbol{k})d_3}\right) \\&\hspace{2cm}-\left( e^{-2q_1(\xi,\boldsymbol{k})d_1}-\Delta_{12}^{\text{TM}}(i\xi,\boldsymbol{k}) \right)\left( e^{-2q_3(\xi,\boldsymbol{k})d_3}-\Delta_{32}^{\text{TM}}(i\xi,\boldsymbol{k}) \right)e^{-2q_2(\xi,\boldsymbol{k})d_2}\Biggr\},
\end{split}
\end{equation*}where
\begin{equation*}\begin{split}
q_j(\xi,\boldsymbol{k})=&\sqrt{\varepsilon_j(i\xi)\mu_j(i\xi)\xi^2+\lambda_{\boldsymbol{k}}^2}=-ip_j(i\xi,\boldsymbol{k}), \hspace{1cm}
\Delta_{jk}^{\text{TM}}(i\xi,\boldsymbol{k})=\frac{r_{jk}^-(i\xi,\boldsymbol{k})}{r_{jk}^+(i\xi,\boldsymbol{k})}=
\frac{\vep_k(i\xi)q_j( \xi,\boldsymbol{k})-\vep_j(i \xi)q_k( \xi,\boldsymbol{k})}{\vep_k(i\xi)q_j( \xi,\boldsymbol{k})+\vep_j(i\xi)q_k( \xi,\boldsymbol{k})}.\end{split}
\end{equation*}Similarly, one finds that
\begin{equation*}
\begin{split}
 f_0(-i\xi, \boldsymbol{k})-f_2(-i\xi,\boldsymbol{k})
=&\frac{\hbar  \xi}{2}e^{ it_c\xi}\frac{d}{d\xi}\log\Biggl\{\left( 1-\frac{r_{12}^-(-i\xi, \boldsymbol{k})}{r_{12}^+(-i\xi, \boldsymbol{k})}e^{-2ip_1(-i\xi, \boldsymbol{k})d_1} \right)\left( 1-\frac{r_{32}^-(-i\xi, \boldsymbol{k})}{r_{32}^+(-i\xi, \boldsymbol{k})} e^{-2ip_3(-i\xi, \boldsymbol{k})d_3}\right)\\&\hspace{2cm}-\left( e^{-2ip_1(-i\xi, \boldsymbol{k})d_1}-\frac{r_{12}^-(-i\xi, \boldsymbol{k})}{r_{12}^+(-i\xi, \boldsymbol{k})}\right)\left(e^{-2ip_3(-i\xi, \boldsymbol{k})d_3}-\frac{r_{32}^-(-i\xi, \boldsymbol{k})}{r_{32}^+(-i\xi, \boldsymbol{k})}\right) e^{-2ip_2(-i\xi, \boldsymbol{k})d_2}\Biggr\}.
\end{split}
\end{equation*}Under the assumption that for all $\xi\in\R$, $j=1,2,3$, $\varepsilon_j(-i\xi)=\varepsilon_j(i\xi)$, $\mu_j(-i\xi)=\mu_j(i\xi)$, we find that
\begin{equation*}
\begin{split}
p_j(-i\xi,\boldsymbol{k})=\sqrt{\varepsilon_j(-i\xi)\mu_j(-i\xi)(-i\xi)^2-\lambda_{\boldsymbol{k}}^2}=
-i\sqrt{\varepsilon_j(i\xi)\mu_j(i\xi)\xi^2+\lambda_{\boldsymbol{k}}^2}=-iq_j(\xi,\boldsymbol{k})
\end{split}
\end{equation*}and
\begin{equation*}
\begin{split}
\frac{r_{jk}^-(-i\xi,\boldsymbol{k})}{r_{jk}^+(-i\xi,\boldsymbol{k})}=&\frac{p_j(-i\xi,\boldsymbol{k})}{p_k(-i\xi,\boldsymbol{k})}\pm\frac{\vep_j(-i\xi)}{\vep_k(-i\xi)}
=
\frac{\vep_k(i\xi)q_j( \xi,\boldsymbol{k})-\vep_j(i \xi)q_k( \xi,\boldsymbol{k})}{\vep_k(i\xi)q_j( \xi,\boldsymbol{k})+\vep_j(i\xi)q_k( \xi,\boldsymbol{k})}=
\Delta_{jk}^{\text{TM}}(i\xi,\boldsymbol{k}).
\end{split}
\end{equation*}Therefore, \begin{equation*}
\begin{split}
 f_0(-i\xi, \boldsymbol{k})-f_2(-i\xi,\boldsymbol{k})
=&\frac{\hbar  \xi}{2}e^{-it_c\xi}\frac{d}{d\xi}\log\Biggl\{\left(1-\Delta_{12}^{\text{TM}}(i\xi,\boldsymbol{k})e^{-2q_1(\xi,\boldsymbol{k})d_1}\right)
\left(1- \Delta_{32}^{\text{TM}}(i\xi,\boldsymbol{k})e^{-2q_3(\xi,\boldsymbol{k})d_3}\right) \\&\hspace{2cm}-\left( e^{-2q_1(\xi,\boldsymbol{k})d_1}-\Delta_{12}^{\text{TM}}(i\xi,\boldsymbol{k}) \right)\left( e^{-2q_3(\xi,\boldsymbol{k})d_3}-\Delta_{32}^{\text{TM}}(i\xi,\boldsymbol{k}) \right)e^{-2q_2(\xi,\boldsymbol{k})d_2}\Biggr\}.
\end{split}
\end{equation*}For the last term in \eqref{eq12_07_1},  under the assumption that   $ \vep_j(\omega)\mu_j(\omega) $ is real for $\omega\geq 0$, we find that if $$\vep_j(\omega)\mu_j(\omega)\omega^2< \lambda_{\boldsymbol{k}}^2,$$ then \begin{equation}\label{eq12_07_2}\begin{split}
\lim_{\epsilon\rightarrow 0^+} p_j\left(\omega+i\epsilon, \boldsymbol{k}\right)=&i\sqrt{\lambda_{\boldsymbol{k}}^2-\vep_j(\omega)\mu_j(\omega)\omega^2},\hspace{1cm}
\lim_{\epsilon\rightarrow 0^+} p_j\left(\omega-i\epsilon, \boldsymbol{k}\right)= -i\sqrt{\lambda_{\boldsymbol{k}}^2-\vep_j(\omega)\mu_j(\omega)\omega^2}
\end{split}\end{equation}whereas if $$ \vep_j(\omega)\mu_j(\omega)\omega^2\geq \lambda_{\boldsymbol{k}}^2,$$ then \begin{equation*}\begin{split}
\lim_{\epsilon\rightarrow 0^+} p_j\left(\omega+i\epsilon, \boldsymbol{k}\right)=&
\lim_{\epsilon\rightarrow 0^+} p_j\left(\omega-i\epsilon, \boldsymbol{k}\right)=  \sqrt{\vep_j(\omega)\mu_j(\omega)\omega^2-\lambda_{\boldsymbol{k}}^2}\geq 0.
\end{split}\end{equation*}Therefore,
\begin{equation*}
\begin{split}
&-\frac{1}{2\pi i}\lim_{\epsilon\rightarrow 0^+}\int_0^{\infty}\left(f_1(\omega+i\epsilon, \boldsymbol{k})-f_2(\omega-i\epsilon,\boldsymbol{k})\right)d\omega\\
=&\frac{\hbar }{4\pi}\lim_{\epsilon\rightarrow 0^+}\int_0^{\infty}\left\{\left(\omega+i\epsilon\right)e^{-t_c(\omega+i\epsilon)}\frac{d}{d\omega}\left(\sum_{j=1}^3 d_j
p_j(\omega+i\epsilon,\boldsymbol{k})\right)+\left(\omega-i\epsilon\right)e^{-t_c(\omega-i\epsilon)}\frac{d}{d\omega}\left(\sum_{j=1}^3 d_j
p_j(\omega-i\epsilon,\boldsymbol{k})\right)\right\}d\omega \\
=&\frac{\hbar}{2\pi}  \sum_{j=1}^3d_j\int_{p_j(\omega,\boldsymbol{k})\geq 0} \omega e^{-t_c\omega} dp_j(\omega,\boldsymbol{k})
\end{split}
\end{equation*}where the part with $ \vep_j(\omega)\mu_j(\omega)\omega^2<\lambda_{\boldsymbol{k}}^2$ has been canceled due to \eqref{eq12_07_2}. Grouping everything together, we have
\begin{equation}\label{eq8_3_7}
\begin{split}
&\frac{\hbar}{2}\sum_{\text{\text{TM} modes}}\omega e^{-t_c\omega}=
 \frac{\hbar}{2\pi} \sum_{\boldsymbol{k}\in\mathcal{S}_{\text{TM}}} \sum_{j=1}^3d_j\int_{p_j(\omega,\boldsymbol{k})\geq 0} \omega e^{-t_c\omega} dp_j(\omega,\boldsymbol{k})\\
&-\frac{\hbar}{2\pi}\sum_{\boldsymbol{k}\in\mathcal{S}_{\text{TM}}}\int_0^{\infty} \xi \cos(t_c\xi) \frac{d}{d\xi}\log\Biggl\{\left(1-\Delta_{12}^{\text{TM}}(i\xi,\boldsymbol{k})e^{-2q_1(\xi,\boldsymbol{k})d_1}\right)
\left(1- \Delta_{32}^{\text{TM}}(i\xi,\boldsymbol{k})e^{-2q_3(\xi,\boldsymbol{k})d_3}\right) \\&-\left( e^{-2q_1(\xi,\boldsymbol{k})d_1}-\Delta_{12}^{\text{TM}}(i\xi,\boldsymbol{k}) \right)\left( e^{-2q_3(\xi,\boldsymbol{k})d_3}-\Delta_{32}^{\text{TM}}(i\xi,\boldsymbol{k}) \right)e^{-2q_2(\xi,\boldsymbol{k})d_2}\Biggr\}d\xi,
\end{split}
\end{equation}

 Notice that the second term in \eqref{eq8_3_7} is finite when $t_c\rightarrow 0^+$. Therefore all the $t_c\rightarrow 0^+$ divergent terms come from the first term in \eqref{eq8_3_7}, which is a linear homogeneous function of $d_1, d_2, d_3$. Setting $t_c=0$ in the second term, we find that when $t_c\rightarrow 0^+$,
\begin{equation}\label{eq8_3_8}\begin{split}
&\frac{\hbar}{2}\sum_{\text{TM modes}}\omega e^{-t_c\omega}\\=&\sum_{j=1}^3d_jX_{j;\text{TM}}(t_c)-\frac{\hbar}{2\pi}\sum_{\boldsymbol{k}\in\mathcal{S}_{\text{TM}}}\int_0^{\infty} \xi \frac{d}{d\xi}\log\Biggl\{\left(1-\Delta_{12}^{\text{TM}}(i\xi,\boldsymbol{k})e^{-2q_1(\xi,\boldsymbol{k})d_1}\right)
\left(1-\Delta_{32}^{\text{TM}}(i\xi,\boldsymbol{k})e^{-2q_3(\xi,\boldsymbol{k})d_3}\right) \\&-\left(  e^{-2q_1(\xi,\boldsymbol{k})d_1}-\Delta_{12}^{\text{TM}}(i\xi,\boldsymbol{k})\right)\left( e^{-2q_3(\xi,\boldsymbol{k})d_3} -\Delta_{32}^{\text{TM}}(i\xi,\boldsymbol{k})\right)e^{-2q_2(\xi,\boldsymbol{k})d_2}\Biggr\}d\xi+o(1),
\end{split}
\end{equation} where $$X_{j;\text{TM}}(t_c)=\frac{\hbar}{2\pi} \sum_{\boldsymbol{k}\in\mathcal{S}_{\text{TM}}} \int_{p_j(\omega,\boldsymbol{k})\geq 0} \omega e^{-t_c\omega} dp_j(\omega,\boldsymbol{k}).$$A comparison of $F_{0;\text{TE}}(\omega,\boldsymbol{k})$ \eqref{eq8_3_10} with $F_{0;\text{TM}}(\omega,\boldsymbol{k})$ \eqref{eq8_3_9} shows that the  expressions for the zero temperature Casimir energy due to the TE modes can be obtained from the expressions for the energy due to the TM modes \eqref{eq8_3_8} by replacing   $\Delta_{jk}^{\text{TM}}$ by $$\Delta_{jk}^{\text{TE}}=\frac{s_{jk}^-(i\xi,\boldsymbol{k})}{s_{jk}^+(i\xi,\boldsymbol{k})}.$$

Next we consider the thermal correction to the Casimir energy
\begin{equation*}
\delta_TE_{\text{Cas}}=k_BT\sum_{\text{modes}}\log\left(1-\exp\left(-\frac{\hbar \omega}{k_B T}\right)\right).
\end{equation*}Define the zeta functions
\begin{equation*}
\zeta_{0}(s)=\sum_{\text{modes}}\omega^{-2s},\hspace{1cm}\zeta_T(s)=\sum_{\text{modes}}\sum_{l=-\infty}^{\infty}\left(\omega^{2}+\left(\frac{2\pi l k_BT}{\hbar}\right)^2\right)^{-s}.
\end{equation*}It is well-known that
\begin{equation*}
\delta_TE_{\text{Cas}}=-\frac{k_BT}{2}\mathcal{Z}'(0),
\end{equation*}where
\begin{equation*}
\mathcal{Z}(s)=\zeta_T(s)-\frac{\hbar}{2\sqrt{\pi}k_BT}\frac{\Gamma\left(s-\frac{1}{2}\right)}{\Gamma(s)}\zeta_0\left(s-\frac{1}{2}\right).
\end{equation*}
Using the generalized Abel-Plana summation formula again, we find that
\begin{equation}\label{eq8_3_2}
\begin{split}
\zeta_{0;\text{TM}}(s)=
 &\frac{1}{\pi} \sum_{\boldsymbol{k}\in\mathcal{S}_{\text{TM}}} \sum_{j=1}^3 d_j\int_{p_j(\omega,\boldsymbol{k})\geq 0} \omega^{-2s}dp_j(\omega,\boldsymbol{k})\\
&+\frac{1}{\pi}\sum_{\boldsymbol{k}\in\mathcal{S}_{\text{TM}}}\int_0^{\infty}\sin\left(\pi s\right)\xi^{-2s} \frac{d}{d\xi}\log\Biggl\{\left(1-\Delta_{12}^{\text{TM}}(i\xi,\boldsymbol{k})e^{-2q_1(\xi,\boldsymbol{k})d_1}\right)
\left(1-\Delta_{32}^{\text{TM}}(i\xi,\boldsymbol{k})e^{-2q_3(\xi,\boldsymbol{k})d_3}\right) \\&-\left(  e^{-2q_1(\xi,\boldsymbol{k})d_1}-\Delta_{12}^{\text{TM}}(i\xi,\boldsymbol{k}) \right)\left( e^{-2q_3(\xi,\boldsymbol{k})d_3} -\Delta_{32}^{\text{TM}}(i\xi,\boldsymbol{k})\right)e^{-2q_2(\xi,\boldsymbol{k})d_2}\Biggr\}d\xi,
\end{split}
\end{equation}
and
\begin{equation}\label{eq8_3_5}
\begin{split}
\zeta_{T;\text{TM}}(s)=
 &\frac{1}{\pi} \sum_{\boldsymbol{k}\in\mathcal{S}_{\text{TM}}} \sum_{l=-\infty}^{\infty}\sum_{j=1}^3 d_j\int_{p_j(\omega,\boldsymbol{k})\geq 0} \left(\omega^2+\left(\frac{2\pi lk_BT}{\hbar}\right)^2\right)^{-s} dp_j\\
&+\frac{1}{\pi}\sum_{\boldsymbol{k}\in\mathcal{S}_{\text{TM}}} \sum_{l=-\infty}^{\infty}\int_{\xi_l}^{\infty}\sin\left(\pi s\right)\left(\xi^{2}-\xi_l^2\right)^{-s} \frac{d}{d\xi}\log\Biggl\{\left(1-\Delta_{12}^{\text{TM}}(i\xi,\boldsymbol{k})e^{-2q_1(\xi,\boldsymbol{k})d_1}\right)
\left(1-\Delta_{32}^{\text{TM}}(i\xi,\boldsymbol{k})e^{-2q_3(\xi,\boldsymbol{k})d_3}\right) \\&-\left(  e^{-2q_1(\xi,\boldsymbol{k})d_1}-\Delta_{12}^{\text{TM}}(i\xi,\boldsymbol{k}) \right)\left( e^{-2q_3(\xi,\boldsymbol{k})d_3} -\Delta_{32}^{\text{TM}}(i\xi,\boldsymbol{k})\right)e^{-2q_2(\xi,\boldsymbol{k})d_2}\Biggr\}d\xi,
\end{split}
\end{equation}where
$$\xi_l=\frac{2\pi |l| k_BT}{\hbar}.$$Therefore, we find from \eqref{eq8_3_2} and \eqref{eq8_3_5} that
\begin{equation*}\begin{split}
&k_BT\sum_{\text{TM modes}}\log\left(1-\exp\left(-\frac{\hbar \omega}{k_B T}\right)\right)=\sum_{j=1}^3 d_j Y_{j;\text{TM}} +\frac{\hbar}{2\pi}\sum_{\boldsymbol{k}\in\mathcal{S}_{\text{TM}}}\int_0^{\infty} \xi \frac{d}{d\xi}\log\Biggl\{\left(1-\Delta_{12}^{\text{TM}}(i\xi,\boldsymbol{k})e^{-2q_1(\xi,\boldsymbol{k})d_1}\right)
\\&\times \left(1-\Delta_{32}^{\text{TM}}(i\xi,\boldsymbol{k})e^{-2q_3(\xi,\boldsymbol{k})d_3}\right) -\left(  e^{-2q_1(\xi,\boldsymbol{k})d_1}-\Delta_{12}^{\text{TM}}(i\xi,\boldsymbol{k}) \right)\left( e^{-2q_3(\xi,\boldsymbol{k})d_3} -\Delta_{32}^{\text{TM}}(i\xi,\boldsymbol{k}) \right)e^{-2q_2(\xi,\boldsymbol{k})d_2}\Biggr\}d\xi\\
&+\frac{k_BT}{2}\sum_{l=-\infty}^{\infty}\log\Biggl\{\left(1-\Delta_{12}^{\text{TM}}(i\xi_l,\boldsymbol{k})e^{-2q_1(\xi_l,\boldsymbol{k})d_1}\right)
\left(1-\Delta_{32}^{\text{TM}}(i\xi_l,\boldsymbol{k})e^{-2q_3(\xi_l,\boldsymbol{k})d_3}\right) -\left(  e^{-2q_1(\xi_l,\boldsymbol{k})d_1}-\Delta_{12}^{\text{TM}}(i\xi_l,\boldsymbol{k}) \right)\\&\times\left( e^{-2q_3(\xi_l,\boldsymbol{k})d_3} -\Delta_{32}^{\text{TM}}(i\xi_l,\boldsymbol{k})\right)e^{-2q_2(\xi_l,\boldsymbol{k})d_2}\Biggr\},
\end{split}\end{equation*}where
\begin{equation*}
\begin{split}
Y_{j;\text{TM}} = \frac{k_BT}{\pi}\sum_{\boldsymbol{k}\in\mathcal{S}_{\text{TM}}}\int_{p_j(\omega,\boldsymbol{k})\geq 0}\log\left(1-\exp\left(-\frac{\hbar\omega}{k_BT}\right)\right) dp_j(\omega,\boldsymbol{k}).
\end{split}
\end{equation*}Together with the zero temperature Casimir energy \eqref{eq8_3_8}, we find that the contribution to the finite temperature Casimir energy from the TM modes is
\begin{equation}\label{eq8_3_11}
\begin{split}
E_{\text{Cas}; \text{TM}}(t_c) =& \sum_{j=1}^3d_j\Xi_{j; \text{TM}}(t_c)+\frac{k_BT}{2}\sum_{\boldsymbol{k}\in\mathcal{S}_{\text{TM}}}\sum_{l=-\infty}^{\infty}\log\Biggl\{\left(1-\Delta_{12}^{\text{TM}}(i\xi_l,\boldsymbol{k})
e^{-2q_1(\xi_l,\boldsymbol{k})d_1}\right)
\left(1-\Delta_{32}^{\text{TM}}(i\xi_l,\boldsymbol{k})e^{-2q_3(\xi_l,\boldsymbol{k})d_3}\right) \\&-\left(  e^{-2q_1(\xi_l,\boldsymbol{k})d_1}-\Delta_{12}^{\text{TM}}(i\xi_l,\boldsymbol{k})\right)\left( e^{-2q_3(\xi_l,\boldsymbol{k})d_3} -\Delta_{32}^{\text{TM}}(i\xi_l,\boldsymbol{k})\right)e^{-2q_2(\xi_l,\boldsymbol{k})d_2}\Biggr\},
\end{split}
\end{equation}where $\Xi_{j;\text{TM}}(t_c)=X_{j;\text{TM}}(t_c)+Y_{j;\text{TM}} $. The contribution to the Casimir energy from the TE modes is obtained from this formula by replacing $\Delta_{jk}^{\text{TM}}$ with $\Delta_{jk}^{\text{TE}}$.

In deriving the formula \eqref{eq8_3_11}, we have made a few assumptions on the nature of the zeros of the functions $F_{0;\text{TM}}(\omega,\boldsymbol{k})$, $F_{0;\text{TE}}(\omega,\boldsymbol{k})$, and the properties of the permittivity and permeability functions $\vep_j(\omega), \mu_j(\omega)$. However,   in general some of these assumptions need not hold. For example, there may be zeros of $F_{0;\text{TM}}(\omega,\boldsymbol{k})$ and $F_{0;\text{TE}}(\omega,\boldsymbol{k})$ that do not give rise to energy eigenmodes. In such cases, we will obtain extra terms for the Casimir energy. Another possible failed assumption is the assumption that $\vep_j(-i\xi)=\vep_j(i\xi)$ and $\mu_j(-i\xi)=\mu_j(i\xi)$. In this case, the second sum in \eqref{eq8_3_11} should be written as half the sum of two terms, one is as shown corresponds to $+i\xi_l$ and the other corresponds to $-i\xi_l$.

\end{document}